\documentclass[10pt,journal,compsoc]{IEEEtran}

\ifCLASSOPTIONcompsoc
  \usepackage[nocompress]{cite}
\else
  \usepackage{cite}
\fi
\ifCLASSINFOpdf
   \usepackage[pdftex]{graphicx}
   \graphicspath{{../pdf/}{../jpeg/}}
   \DeclareGraphicsExtensions{.pdf,.jpeg,.png}
\else

\fi

\usepackage{amsmath}

\usepackage{algorithmic}

\usepackage{array}

\usepackage{xcolor}
\usepackage{amsmath}
\usepackage{amsthm}
\usepackage{amssymb}
\usepackage{subfigure}
\usepackage{xcolor,graphicx,float}
\usepackage{multirow}
\usepackage{longtable}
\usepackage{booktabs}
\usepackage{makecell}
\usepackage{amsfonts}
\usepackage{amssymb}
\usepackage{bbm}
\usepackage{bm}
\begin{document}

\title{CrowdOS: A Ubiquitous Operating System for Crowdsourcing and Mobile Crowd Sensing}

\author{Yimeng~Liu,~\IEEEmembership{Student~Member,~IEEE,}
        Zhiwen~Yu,~\IEEEmembership{Senior~Member,~IEEE,}
        Bin~Guo,~\IEEEmembership{Senior~Member,~IEEE,}
        Qi~Han,~\IEEEmembership{Senior~Member,~IEEE,}
        Jiangbin~Su,
        Jiahao~Liao
\IEEEcompsocitemizethanks{\IEEEcompsocthanksitem Y. Liu, Z. Yu, B. Guo, J. Su, J. Liao are with the Department
of Computer Science, Northwestern Polytechnical University, Xi'an 710129, China.  \protect\\
E-mail: {flash\_lym@mail.npu.edu.cn; zhiwenyu@nwpu.edu.cn; guobin.keio@gmail.com; \{sujb0808, liaojh\}@mail.nwpu.edu.cn}
\IEEEcompsocthanksitem Q. Han is with the Department of Computer Science, Colorado School of Mines, Golden, CO 80401, USA.   \protect\\
E-mail: {qhan@mines.edu}

}
\thanks{}}

\markboth{}%
{Shell \MakeLowercase{\textit{et al.}}: Bare Demo of IEEEtran.cls for Computer Society Journals}

\IEEEtitleabstractindextext{%
\begin{abstract}
With the rise of crowdsourcing and mobile crowdsensing techniques, a large number of crowdsourcing applications or platforms ($\mathbb{CAP}$) have appeared. In the mean time, $\mathbb{CAP}$-related models and frameworks based on different research hypotheses are rapidly emerging, and they usually address specific issues from a certain perspective. Due to  different settings and conditions,  different models are not compatible with each other. However, $\mathbb{CAP}$ urgently needs to combine these techniques to form a unified framework. In addition, these models needs to be learned and updated online with the extension of crowdsourced data and task types, thus requiring a unified architecture that integrates lifelong learning concepts and breaks down the barriers between different modules. This paper draws on the idea of ubiquitous operating systems and proposes a novel OS (CrowdOS), which is an abstract software layer running between native OS and application layer. In particular, based on an in-depth analysis of the complex crowd environment and diverse characteristics of heterogeneous tasks, we construct the OS kernel and three core frameworks including Task Resolution and Assignment Framework (\emph{TRAF}), Integrated Resource Management (\emph{IRM}), and Task Result quality Optimization (\emph{TRO}). In addition, we validate the usability of CrowdOS, module correctness and development efficiency. Our evaluation further reveals \emph{TRO} brings enormous improvement in efficiency and a reduction in energy consumption.

\end{abstract}

}

\maketitle

\IEEEdisplaynontitleabstractindextext

\IEEEpeerreviewmaketitle

\IEEEraisesectionheading{\section{Introduction}\label{sec:introduction}}

\IEEEPARstart{T}{he} concept of crowdsourcing \cite{howe2006rise} can be traced back to 2006. The basic idea is to divide one complicated or massive task into multiple smaller tasks that can be completed by multiple participants. Crowdsourcing has also been used to find the best solution provider: participants are responsible for providing the solutions, while publishers pay participants for each solution.

With the advent and evolution of Web 2.0 technologies, crowd intelligence has beed adopted as a powerful method to solve social problems. Many crowdsourcing platforms,  have emerged to solve general problems. Examples include Amazon Mechanical Turk \cite{buhrmester2011amazon}, CrowdFlower\cite{van2012designing}, a platform for promoting gastronomic tourism \cite{michail2019bucketfood}, and a platform for intelligence analysis \cite{xia2019trace}.  Many crowdsourcing applications \cite{lopez2010peoplecloud}, \cite{sabou2012crowdsourcing}, \cite{alt2010location} have also been developed to solve specific problems.

We consider these applications and platforms as the first generation of crowdsourcing technology. This generation shares several common characteristics: relying on the web intermediary platform to publish heterogeneous tasks and collect results, applying the principle of `divide and conquer' for large-scale problems. This generation does not focus on the task itself, nor does it optimize results quality.

In recent years, complex, fixed, and non-programmable sensors have evolved  to being small, portable, and programmable. Smart devices such as mobile phones, tablets, wristbands, watches are also embedded with multiple sensors, which makes it easier for people to access various sensor data. These advances in combination with crowd intelligence make it possible for us to obtain massive amount of heterogeneous sensor data to solve multi-domain problems. Many   Mobile CrowdSensing (MCS) \cite{ganti2011mobile}, \cite{guo2015mobile} applications have been developed for environmental monitoring, public facilities monitoring, and social networking. Examples include Common Sense \cite{dutta2009common}, Ear-Phone \cite{rana2010ear}, Chimera \cite{pu2018chimera}, Creekwatch \cite{kim2011creek}, and PhotoCity \cite{tuite2011photocity}. Various research topics in MCS have  also been investigated. For example, Wang et al. studied the problem of multi-task cooperative allocation in MCS applications \cite{guo2016activecrowd}, \cite{wang2018heterogeneous}. Guo et al. proposed challenges for the optimization of sensing data. F.~Restuccia et al. studied the quality improvement of crowdsensing data \cite{cheng2015deco}, \cite{restuccia2018first}.  In addition, incentive mechanisms for crowdsourcing workers and privacy protection has been a well-studied topic \cite{xie2016incentive}.

We consider these applications and studies as the second generation of crowdsourcing technology. These applications are domain- and task-specific, so the software design does not consider reusability, scalability, and portability. Most research here is based on strong assumptions that only hold in an ideal environment.  Often only simulations have been conducted to verify the proposed ideas. Design rules and parameter sizes are all different, so these studies are isolated from each other and  insights obtained from these studies cannot be easily applied to practical applications. These have significantly limited a wide adoption of crowdsourcing technologies.

Based on our comprehensive theoretical analysis and practical experience with the first and second generations of crowdsourcing technologies, we have identified the following major challenges in developing $\mathbb{CAP}$.

\begin{enumerate}
\item $\mathbb{CAP}$ are task driven, so $\mathbb{CAP}$ similar to bulletin boards that only publish tasks and collects results are not powerful enough. There is an urgent need for a framework that can seamlessly integrate tasks and platforms while handling various homogeneous and heterogeneous tasks. 

\item Existing $\mathbb{CAP}$ are more concerned with the implementation of functionalities, so they rarely pay special attention to system resource management. Due to a diverse range of hypothesis and scene settings, it is difficult to conduct research on holistic resource management.

\item Current $\mathbb{CAP}$ only summarize task results without efficiently assessing and improving the quality of task results. Although result filtering methods may be embedded in $\mathbb{CAP}$ to improve data quality, these methods are usually only for specific tasks and data types, and cannot be applied to other tasks. 

\item Most of the current technical research is to solve specific problems in an ideal environment, along with many assumptions. Since these studies are scattered and isolated from each other, it is difficult to apply and promote these methods.
\end{enumerate}

As we try to design systematic approaches to address these issues, operating system (OS) design ideas come into our view. As the kernel and cornerstone of a computer, an OS uniformly manages system resources. Different types of traditional OS have different focus in their design: Servers often use Linux and Unix that can be tailored to user needs; Windows and Mac OS pay more attention to GUI; Android, iOS and Windows Phone often use mobile OS with rich components and lightweight libraries to support mobile app development.

In the future, operating systems will be ubiquitous \cite{mei2018toward}. In fact, a number of ubiquitous operating systems have already emerged. For example, TinyOS is an OS for wireless sensor networks \cite{levis2005tinyos}, and ROS is an Open-Source robot OS \cite{quigley2009ros}. There are also operating systems for Home \cite{dixon2012operating}, Campus \cite{yuan2013towards}, and Internetware \cite{mei2012internetware}. These micro OS usually exist on the upper layer of the native OS. They not only manage heterogeneous hardware resources in the system, but also provide unique resource abstraction and software-defined services for different application scenarios. They are a higher level OS, providing a wealth of functional components and software development kits. Their existence are mainly due to a large number of application scenario and need for various features.

Building on these rapidly evolving technologies and ideas, we have designed CrowdOS, a novel ubiquitous operating system for crowdsourcing and MCS. This article presents the core architecture and design principles of CrowdOS and discusses how CrowdOS addresses the aforementioned challenges. Specifically, we make the following contributions.

\begin{itemize}
\item We design the core architecture of CrowdOS to tackle challenge 4), and provide a unified definition and workflow of $\mathbb{CAP}$ tasks (Section 3). Unlike existing ubiquitous OS or frameworks, CrowdOS is the first work that can deal with multiple types of crowdsourcing problems simultaneously.

\item To address challenge 1), we propose a Task Resolution and Assignment Framework (\emph{TRAF}) that can understand and memorize the important characteristics of various tasks like humans. By exploiting rich semantic information and discrete features, we construct a fine-grained vector for each task. Resource scheduling and task assignment are then implemented with the collective support of task resource graphs and assignment strategies (Section 4).

\item We design an Integrated Resource Management (\emph{IRM}) framework to deal with challenge 2), and conduct a thorough analysis to implement virtual and physical entities, heterogeneous multimodal data, and knowledge base management. (\emph{IRM}) is a service oriented management paradigm to support a family of methods and models.

\item To address challenge 3), we propose a Deep Feedback Framework based on Human-Machine Interaction (\emph{DFHMI}) (Section 6). A quality assessment mechanism and a shallow-deep inference mechanism are designed to uniformly support the implementation of strategies for different quality issues.
\end{itemize} 

In addition, Section 2 reviews previous work related to this paper. Section 7 briefly describes the components, libraries and interfaces of CrowdOS. Section 8 presents the evaluation and results. Section 9 discusses the current limitations of CrowdOS and future work. We conclude the paper in Section 10.

\section{Related Work}
The representative work from two main aspects is discussed: crowdsourcing-related frameworks and ubiquitous operating systems.

\subsection{Framework for CAP Problems}
There have been many frameworks designed to address $\mathbb{CAP}$-related issues.

\emph{Task Assignment Framework.} In \cite{cheng2018frog}, Cheng et al. proposed FROG, which consists of task scheduler and notification modules and assigns tasks to suitable workers with high reliability and low latency. The approaches such as request-based, batch-based, and smooth kernel density estimation was used. Alireza et al. in \cite{moayedikia2018task} introduced an algorithm LEATask with two stages of exploration and exploitation. It assigns tasks to new workers by assessing the similarities in performance of workers. However, the algorithm needs to hire some workers in the early stage to learn their reliability and cluster them. Wang et al. in \cite{wang2017psallocator} exploit task allocation framework in participatory sensing. Based on the prediction of the connection of the participant to the cellular tower and the location obtained by historical data from the telecom operator, an iterative greedy process is employed to optimize the task allocation. Because most framework implementations need to collect relevant data in advance and complete a series of steps or have specific domain datasets, it is difficult for the framework to be extended to general or new types of tasks and scenarios.

\emph{Crowdsourcing Resource Management Framework.} In \cite{atzori2017siot}, Atzori et al. built an MCS management framework on top of the social IoT lysis platform, where social virtual objects resources are fairly allocated so that no node are overloaded. A resource optimization method in \cite{rizvi2015mediaserv} was designed for content delivery, using some discrete time slots or transmission opportunities to deliver media contents to the service points when the network connectivity is intermittent. In addition, Meng et al. proposed an optimal real-time pricing strategy for computer resource management in \cite{meng2017optimal}, where computing resources are managed to benefit the overall system. However, these works mainly focus on single aspect of system resource or management of homogeneous device, while our paper proposes an integrated management framework, from the perspective of heterogeneous equipment management, multi-type resource management and so on.

\emph{Quality Optimization Framework.} Fabio et al. evaluated 76 crowdsourcing projects found in 72 articles in \cite{neto2018understanding}, which helped researchers and crowdsourcers to understand the state-of-the-art of crowdsourcing and evaluated the quality management in crowdsourcing projects preliminary. In \cite{alabduljabbar2019dynamic}, Reham et al. proposed a dynamic approach for selecting the best quality control mechanism for a task rather than selecting a special one for all types of tasks. Oleson et al. present an inexpensive and scalable automated quality assurance process in \cite{oleson2011programmatic}, which relies on programmatic gold creation to provide targeted training feedback to workers and to prevent common scamming scenarios. This reduces the amount of manual work required to manage crowdsourced labor while improving the overall quality of the results. Nevertheless, our \emph{DFHMI} not only customize the correction strategy and operations for each unqualified task in real time, but also can dynamically expand the strategy library and operations.

\emph{Other Frameworks.} In addition to the above specific framework for specific issues, there are some other broader frameworks. In \cite{diniz2016reference}, with the intention of offer a straightforward and easy-to-follow reference architecture, Herbertt et al. present an approach that employs off-the-shelf components for the construction of an MCS platform for participatory sensing solutions in Smart Cities and demonstrate the architecture in a specific domain. In \cite{fan2015icrowd}, Fan et al. presents an adaptive crowdsourcing framework, iCrowd. It on-the-fly estimates accuracies of a worker by evaluating her performance on the completed tasks, and predicts which tasks the worker is well acquainted with, thereby improving subsequent task assignments.

\subsection{Ubiquitous Operating System}
This section mainly introduces several UOS for different fields and scenarios. They all have their own unique design perspectives and  a systematic approach to solving domain problems. In the end, we summarize the commonality and uniqueness of CrowdOS compared to these UOS.

HomeOS \cite{dixon2012operating} is a platform that simplifies the task of managing and extending technology in the home by providing a PC-like abstraction for network devices to users and developers. It uses network devices as peripherals with abstract interfaces, implements cross-device tasks through applications written for these interfaces, and provides users with a management interface designed for the home environment. HomeOS already has dozens of applications and supports a variety of devices.

CampusOS \cite{yuan2013towards} is an operating system that manages the network resources of a university campus. It provides flexible support for campus application development through an SDK that includes campus-related APIs. Developers can also easily extend the OS feature and the SDK.

Terrence et al. are developing an interaction infrastructure called the Human-Robot Interaction Operating System (HRI/OS) \cite{fong2006human}. The HRI/OS provides a structured software framework for building human-robot teams, supporting a variety of user interfaces, enabling humans and robots to participate in task-oriented conversations and facilitating robot integration through extensible APIs.

ROS is an open source robot operating system. It relies on native OS of heterogeneous computing clusters and provides a structured communication layer on top of them. In \cite{quigley2009ros}, they discussed how ROS is associated with existing robotic software frameworks and provides a brief overview of several available applications that use ROS.

Additionally, Urban OS \cite{urbanOS} was proposed as a software platform to accelerate urban technology development and equipment deployment. While, BOSS \cite{dawson2013boss} provides a set of system services to support applications deployed on distributed physical resources in large commercial buildings.

It can be seen from the analysis that these operating systems are designed to solve general problems in their field, and provide a comprehensive framework and a rich set of APIs. To the best of our knowledge, this article is the first work to explore the principles and architecture of the crowd operating system to address $\mathbb{CAP}$-related issues. We have designed and implemented the overall architecture, important mechanisms and functional components of CrowdOS, while also taken into account the internal interaction and external callable interface, as well as the stability and scalability of the system.


\section{Task Definition and System Architecture}
In this section, we present a unified definition of a crowdsourcing task and its execution process and phases, describe the architecture of CrowdOS and the relationships between different modules, and explain system resource graph.

\begin{figure}[!tp]
\setlength{\abovecaptionskip}{-0.0cm}
\setlength{\belowcaptionskip}{0.5cm}
\centering
\includegraphics[width=3.5in]{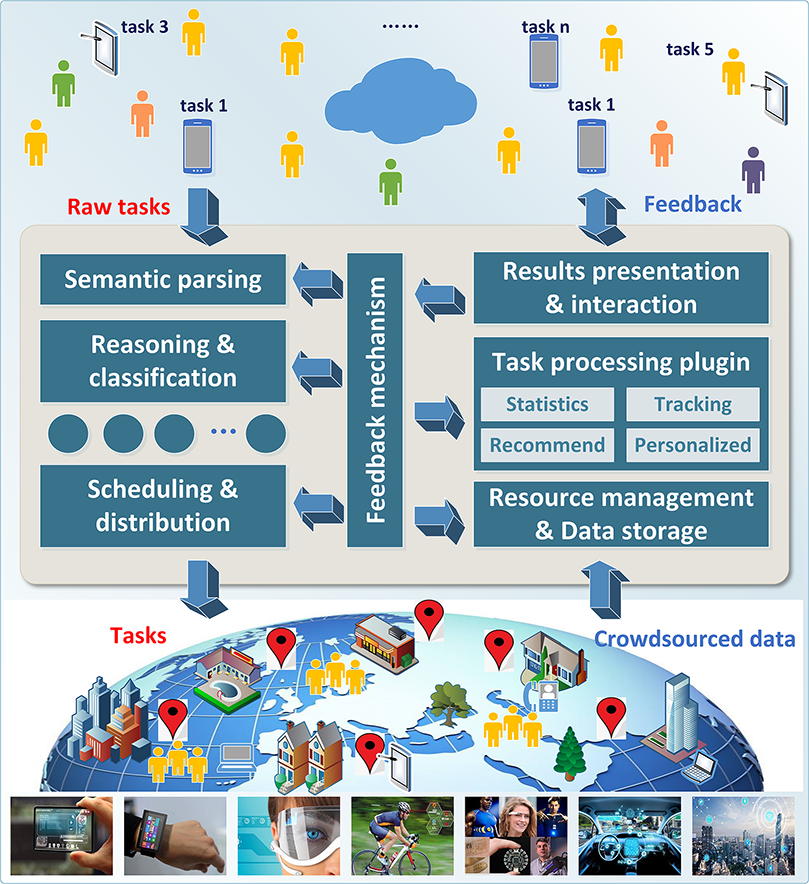}
\caption{Crowdsourcing ecosystem framework.}
\label{fig1}
\end{figure}

\subsection{Definition of Crowdsourcing Tasks }
We present a formal definition of homogeneous and heterogeneous tasks in crowdsourcing and MCS, in which participants (either human workers or sensing devices) are selected to perform tasks within crowdsourcing ecosystem as shown in Fig. \ref{fig1}. For ease of the following presentation, we list the notations used frequently in Table \ref{table1}.

\begin{table}[bh]\footnotesize
\newcommand{\tabincell}[2]{\begin{tabular}{@{}#1@{}}#2\end{tabular}}
\caption{Frequently Used Notations}%
\label{table1}
\centering
 \vspace{-0.25cm} 
\begin{tabular}{p{2.0cm}p{5.7cm}}
\toprule
\multicolumn{1}{l}{\textbf{Symbol}} & {\textbf{Meaning}}\\
\midrule
\tabincell{l}{\emph{$\mathbb{T}$}$\{t_{i\in{N}}\}$} & \tabincell{l}{Crowdsourcing and MCS tasks}\\

\tabincell{l}{$TID_{t_i}(id)$} & \tabincell{l}{Task identifier}\\

\tabincell{l}{$U({p_{t_i}^h, w_{t_i}^j, d_{t_i}^k})$} & \tabincell{l}{Publishers, participant workers, device}\\

\tabincell{l}{$TF_{t_i}=\{clf_{t_i}^m\}$} & \tabincell{l}{Task classification, class $m$}\\

\tabincell{l}{$TD_{t_i}(l,dsp)$} & \tabincell{l}{Language, task detailed description}\\

\tabincell{l}{$TC_{t_i}(t,s,o)$} & \tabincell{l}{Task termination condition}\\

\tabincell{l}{$TP_{t_i}=\{ph_{t_i}^m\}$} & \tabincell{l}{$t_i$ task phases, phase $m$}\\

\tabincell{l}{$QAM=\{Q_{t_i},\delta\}$} & \tabincell{l}{Result quality, quality threshold}\\

\tabincell{l}{$\mathbb{CES}\{\epsilon_{i\in{N}}^{k_i}\}$} & \tabincell{l}{Crowdsourcing ecosystem, $i$-types, $k$-elements} \\
\bottomrule
\end{tabular}
\end{table}

We first define crowdsourcing tasks.

\textbf{Definition 1 (Heterogeneous crowdsourcing tasks).} \emph{$\mathbb{T}$}$\{t_{i\in{N}}\}$ is a set of crowdsourcing tasks. Each $t_i$ corresponds to a five-element tuple $\{TID_{t_i},U_{t_i},TF_{t_i},TD_{t_i},TC_{t_i}\}$, which presents the important characteristics of $t_i$. The first four elements denote customized identifier, participating entities, types and detailed description of $t_i$, and $TC_{t_i}(t,s,o)$ contains the task termination condition.

Current crowdsourcing application is mainly oriented to the needs of social, institutional or individual. Therefore, task publisher $p_{t_i}^h$ could be individual, organization, or government agency and participants include workers $w_{t_i}^j$ and automated sensing equipment $d_{t_i}^k$. The parameters $t, s, o$ of $TC_{t_i}$ indicate that $t_i$ will be terminated when a certain time, scale or additional conditions are satisfied.

The survival and execution of tasks depends on the crowdsourcing ecosystem. The crowdsourcing ecosystem $\mathbb{CES}$ is also a typical Cyber-Physical-Social Systems. From the bottom up, it contains $n$ elements such as server clusters $\epsilon_1$, smart terminals $\epsilon_2$, mobile and fixed sensing devices $\epsilon_3$, native OS, communication networks $\epsilon_i$, basic software layers, uOS, platforms and applications, publishers $\epsilon_{n-1}$ and participants $\epsilon_n$.

\textbf{Definition 2 (\emph{$\mathbb{CAP}$} Problems).} In $\mathbb{CES}$, applications and platforms are defined as $\mathbb{CAP}$, through which \emph{$\mathbb{T}$} are released and executed. A typical $\mathbb{CAP}$ includes a variety of crowdsourcing, crowdsensing, mobile crowdsourcing, and mobile crowd sensing applications or platforms. Therefore, the $\mathbb{CAP}$ problem refers to a series of related problems in the process of perception, analysis, calculation, and management performed on $\mathbb{CAP}$.

In Fig. \ref{fig1}, a task transfer process can be divided into three parts. In the release process, users publish tasks through a special \emph{CAP}, \emph{Cap1}. In the implementation process, \emph{Cap1} brings together participants who meet tasks execution conditions and contribute their terminal sensing ability, device computing power and human intelligence. In the feedback process, If the publisher is not satisfied with task results $Q(t_i)<\delta$, system or workers need to further correct result and evaluate again until it is qualified $Q(t_i)\ge\delta$.

The detailed dynamic execution phase of $t_i$, $TP_{t_i}$, is defined as follows.

\textbf{Definition 3 (Task Execution Phase).} Given crowdsourcing task $t_i$, the deduction rules between the current execution phase and the next phase is formulated as follows:

\begin{small}
\begin{equation} \label{eq1}
TP_{t_i}=\begin{cases}
ph_{t_i}^k \to ph_{t_i}^{k+1},&0<k<{\Gamma-1} \\
ph_{t_i}^{k=\Gamma} \to ph_{t_i}^{l\in\theta},&Q(t_i)<\delta \\
end,  & Q(t_i)\ge\delta
\end{cases}
\end{equation}
\end{small}

As Eq. (\ref{eq1}) shows, we divide the task execution process $TP_{t_i}$ into several phases $ph_{t_i}^k$. $\Gamma$ is the total number of phases. $\theta$ is a set of phases that can be reached through feedback mechanism. Fig. \ref{fig1} shows the overall execution process of a task and relationships between the phases. The content of each phase is as follows:

$\bm{ph_{t_i}^1}$: Creation phase. Task publishers input raw tasks through terminals such as smartphones and submit them to \emph{Cap1}. It captures new tasks and assign unique task identifiers $TID_{t_i}$ to each $t_i$, which will accompany their entire lifecycle.

$\bm{ph_{t_i}^2}$: Generation phase. \emph{Cap1} performs task analysis and generates corresponding task feature vector. Through the task vector, \emph{Cap1} can extract important characteristics such as task type, participants size, location, required sensors, etc.

$\bm{ph_{t_i}^3}$: Assignment phase. \emph{Cap1} completes user scheduling and task assignment process by performing task analysis, resolution, strategy selection, and related operations.

$\bm{ph_{t_i}^4}$: Execution phase. Participants who have received the task upload their collected sensor data or design documents to \emph{Cap1}. \emph{Cap1} classifies and stores these heterogeneous multimodal data.

$\bm{ph_{t_i}^5}$: Processing phase. According to detailed task features description, \emph{Cap1} selects middlewares and summarizes the collected data. Using multiple types of processing plug-ins to complete data statistics, information mining, etc.

$\bm{ph_{t_i}^6}$: Feedback phase. The publisher evaluates the final results quality to determine whether \emph{Cap1} needs to further refine the result based on the feedback information.

$\bm{ph_{t_i}^7}$: Termination phase. \emph{Cap1} presents the results to the publisher in a standard format, with the attached data in the download link.

A large number of tasks are executed in \emph{Cap1}. In order to maximize system resource utilization, \emph{Cap1} abstracts tasks and software defines tasks, users and other resources, then uniformly schedules and manages them. The cycle of a task $t_i$ begins with the user editing the task $ph_{t_i}^1$ and ends when the user provides positive feedback to the task results $ph_{t_i}^7$.

\subsection{CrowdOS Kernel}

\begin{figure}[!t]
\centering
\setlength{\abovecaptionskip}{0.cm}
\setlength{\belowcaptionskip}{-0.cm}
\includegraphics[width=3.5in]{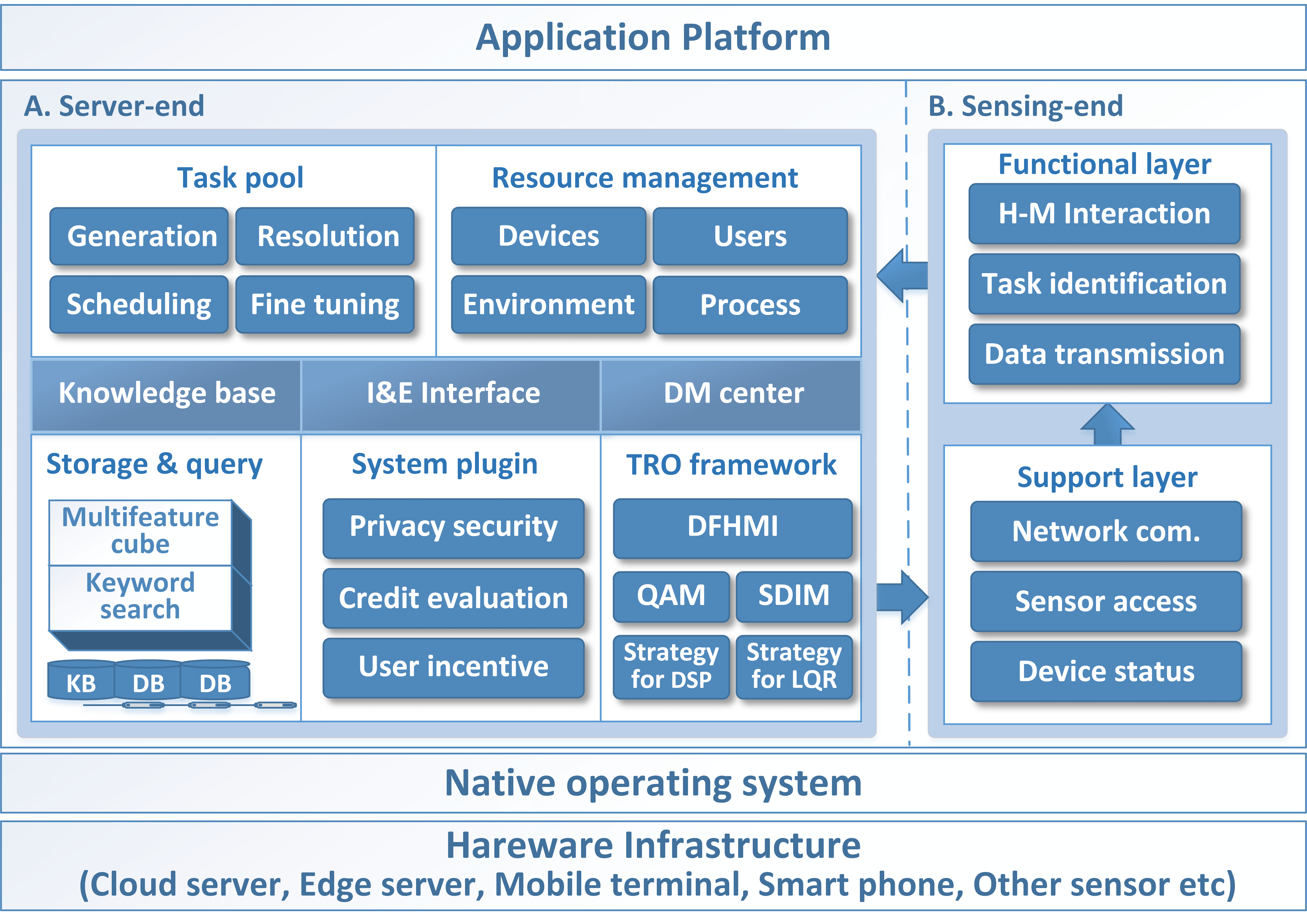}
\caption{CrowdOS architecture.}
\label{fig2}
\end{figure}

Based on the analysis of $\mathbb{CAP}$ problems, we propose the Crowd Operation System (CrowdOS). It is designed to comprehensively address versatility and isolation $\mathbb{CAP}$ problems. The OS kernel architecture is presented in Fig.~\ref{fig2} and we introduce the relationships between various modules. CrowdOS not only shields the differences of native OS running on heterogeneous devices, but also reserves interfaces of extensible function modules and personalized plugins.

CrowdOS runs between the native operating system and the upper application. It includes the sensing-end and server-end. Sensing-end software  consists of two types of devices. The first type is portable smart sensing devices with human-machine interaction functionalities, such as smart phones and smart watches. The second type is fixed sensors deployed in the physical world which do not need to interact with people directly, such as vehicle sensors, water quality sensors, air quality sensors. Server-end software provides integrated management services, which are usually deployed on server clusters, cloud servers, or edge servers. The core processing mechanisms of the OS, such as task assignment and scheduling, resource storage and management, are deployed in server end. Two ends perform data transfer and behavior control through a set of communication and interaction protocols we define.

{\it Sensing-end} is divided into two layers. The bottom layer is the system support layer, which is mainly responsible for the following functions:

\begin{itemize}
\item Get device status, such as current device availability, remaining power, location, etc.
\item Unify packaging of sensor interfaces and data transfer formats.
\item Capture available communication types and modes of device , then store them in structures.
\end{itemize}

The upper layer is the functional layer, which mainly completes three types of operations: human-machine (\emph{H-M}) interaction, task identification, and data transmission. Raw tasks can be uploaded to servers by publisher through the \emph{H-M} interactive module. Participants can browse and execute tasks that have been published through smart terminals. However, for fixed sensors without the \emph{H-M} interaction module, once they are activated by the authenticated tasks, they automatically collect and upload sensor data according to predetermined rules.

{\it Server-end} combines multiple modules and innovative mechanisms. It is mainly responsible for task scheduling and assignment, resource storage, and management, data processing, and result optimization. Server-end not only handles tasks in a fine-grained manner, but also builds a unified knowledge base, while also providing a rich set of crowdsourcing components as system plugins. Next, we briefly introduce the function of each module.

\begin{itemize}
\item Task pool module performs operations such as parsing, scheduling, allocation, and fine-tuning on eceived raw tasks.
\item Resource Management module comprehensively manages the heterogeneous sensing devices, environment resources, users and task process.
\item Storage and query module provides categorized storage and rapid retrieval of massive amounts of heterogeneous data.
\item System plugin module provides a wealth of crowdsourcing components such as privacy protection, security, credit evaluation, and user incentives.
\item Task Result Optimization (\emph{TRO}) framework is primarily designed to optimize results quality, which consists of Deep feedback Framework based on Human-machine Interaction (\emph{DFHMI}), Quality Assessment Mechanism (\emph{QAM}), Shallow-Deep Inference Mechanism (\emph{SDIM}) and specific strategies.
\item Data Management Center (\emph{DMC}) is mainly responsible for managing data that come with tasks, uploaded by participants, or generated during the execution process. Most of these heterogeneous multi-source multimodal data are unstructured. Knowledge Base (\emph{KB}) is the basis and premise of system to make reasoning. Constructing \emph{KB} is an efficient way to systemically manage domain knowledge. Internal and External (\emph{I-E}) interfaces include system internal interface and CrowdAPI, where the internal interface is a set of protocols used for system testing and interaction between modules. CrowdAPI provides a unified call interface for application development.
\end{itemize}

CrowdOS is designed using cloud-edge-side architecture: sensing-end is deployed on terminals to collect sensing data and special task solutions; server-end is deployed on a cloud or edge servers, which is responsible for comprehensive management of resources and real-time response to system operations; when deployed on edge servers, the OS is usually tailored and lightweight.

\subsection{System Resource Graph}

CrowdOS abstracts and defines various entities and virtual resources in $\mathbb{CES}$. By constructing five dynamic agents to generate System Resource Graph (\emph{SRG}), tasks and resources in the system are managed in a unified manner. Five agents include \emph{Task-Agent} (\emph{TA}), \emph{User-Agent} (\emph{UA}), \emph{Device-Agent} (\emph{DA}), \emph{Environment-Agent} (\emph{EA}), and \emph{Process-Agent} (\emph{PA}).

\begin{figure}[!htp]
\setlength{\abovecaptionskip}{0.cm}
\setlength{\belowcaptionskip}{-0.cm}
\centering
\includegraphics[width=3in]{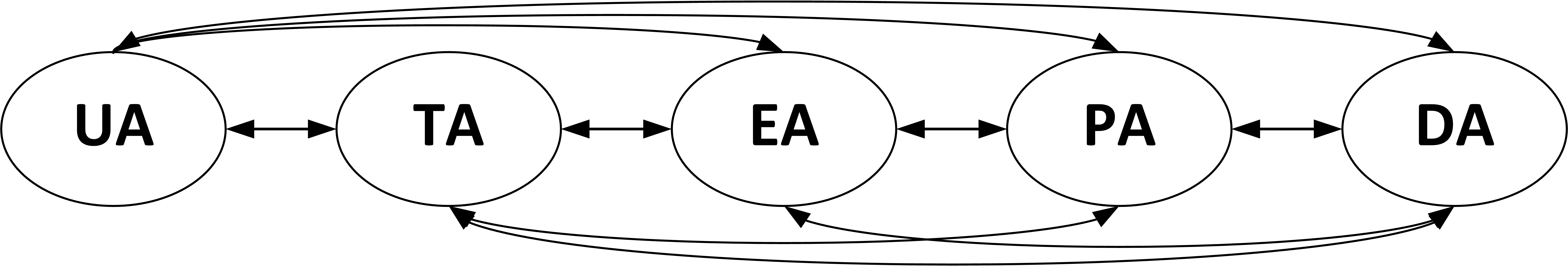}
\caption{System Resource Graph.}
\label{fig3}
\end{figure}

As shown in Fig. \ref{fig3}, agents communicate with each other, abstracting and defining all the resources in system. \emph{TA} contains detailed information about each crowdsourcing task, which is the parsing and perfection of the five-element tuple in definition 1. \emph{UA} is an abstraction of user, which records relevant information such as published and executed tasks, credit ratings or interests. \emph{DA} is a description of the terminal or sensor devices, recording information such as type specifications and current states. \emph{EA} abstracts the hardware and software environment resources in current \emph{CAP}, including CPU utilization, remaining memory or storage capacity, as well as user volume and the total number of available devices. \emph{PA} manages task processes in \emph{CAP}, including process status, priority, scheduling policy, etc.  These agents are used by other modules in the architecture.

\section{Task Resolution and Assignment Framework}

This section focuses on task resolution, user scheduling, and task assignment process in \emph{CAP}. The process begins with publishers editing and submitting tasks, continuing until tasks are assigned to appropriate participants. As one of the core components in OS kernel, it attempts to solve the first challenge: adaptively handling multiple types of crowdsourcing tasks uniformly.

To address the challenge, there are two key points to emphasize. First, tasks need to be analyzed in a fine-grained manner and deeply understood in order to extract their commonalities and differences. Second, a reasonable allocation strategy needs to be chosen to ensure that tasks are completed in shortest time or lowest energy consumption. We next suggest the best operating method in realizing the multi-task resolution and adaptive task assignment.

\begin{figure}[!htp]
\centering
\setlength{\abovecaptionskip}{0.cm}
\setlength{\belowcaptionskip}{-0.cm}
\includegraphics[width=3.5in]{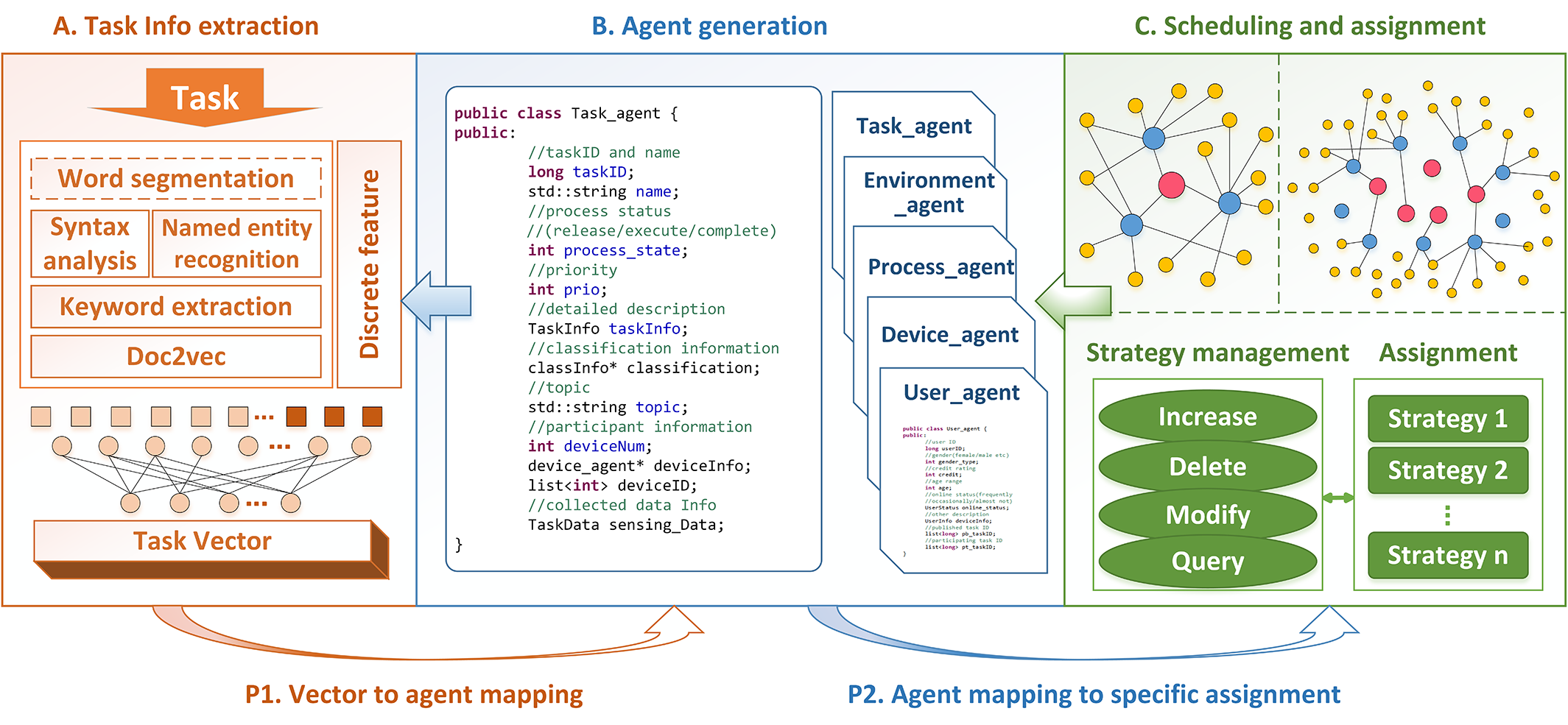}
\caption{Task resolution and assignment framework.}
\label{fig4}
\end{figure}

\subsection{Task Resolution based on Semantic Analysis}
Crowdsourcing tasks differs from typical regular tasks in at least two aspects. The first is different patterns of thinking and different customs of language expression. The same problem can be described in a variety of ways. The second is the ambiguity of task results. For the same task, there is an infinite combination of execution results that satisfy the condition due to the differences in participants or differences in execution time. We next focus our discussion on how to addresses the diversity of natural language description of crowdsourcing tasks. More specifically, we discuss unified coding of information about tasks and abstracting them into task agents, ultimately sharing tasks in a machine-understandable way.

Fig. \ref{fig4}-A shows the process of semantic parsing and feature extraction. The system performs natural language analysis on the received tasks. For tasks described in Chinese, Japanese, etc., the system performs word segmentation first. The operations of part-of-speech tagging, named entity recognition, and keyword extraction will be performed. Finally we extract task-critical information, such as the way to perform the task, location, time, number of participants. The system further connects the extracted task-critical information and the discrete features  obtained by the button clicking or rules selection. The stitched features are fed into a deep neural network for unified encoding, which outputs a high-dimensional intermediate vector (task vector). Finally, the vector is mapped to \emph{TA} by decoding, thus the conversion process \emph{P1} in Fig. \ref{fig4} is completed.

The reduced structure of \emph{TA} is shown in Fig. \ref{fig4}-B. \emph{TA} contains all the common and individual information of the task. \emph{TaskID} is the unique identifier of the task in the system. \emph{Process-state} indicates the current state of the task process, including the generated state, the execution state, or the feedback state, etc. This state will assist \emph{Process-agent} in task process management, which will be described later. \emph{Prio} represents the priority of the task, from 0-15. System schedules the task process according to the priority order. \emph{taskInfo} is a structure that contains task details such as the execution time range and location, vector representation. Classification represents the category to which the task belongs, such as data annotation class, sensor information collection class, and questionnaire investigation class. Topic shows the theme of the task, which can be extracted from  keywords, such as audio collection, photo collection. The \emph{deviceNum}, \emph{deviceInfo}, and \emph{deviceID} represent the number of available devices, device details, and device IDs. \emph{Sensing Data} is the pointer to the buffer that contains the cube address where collected data is stored.

\subsection{Resource Scheduling and Task Assignment}
\begin{figure}[!htp]
\centering
\setlength{\abovecaptionskip}{0.cm}
\setlength{\belowcaptionskip}{-0.cm}
\includegraphics[width=3in]{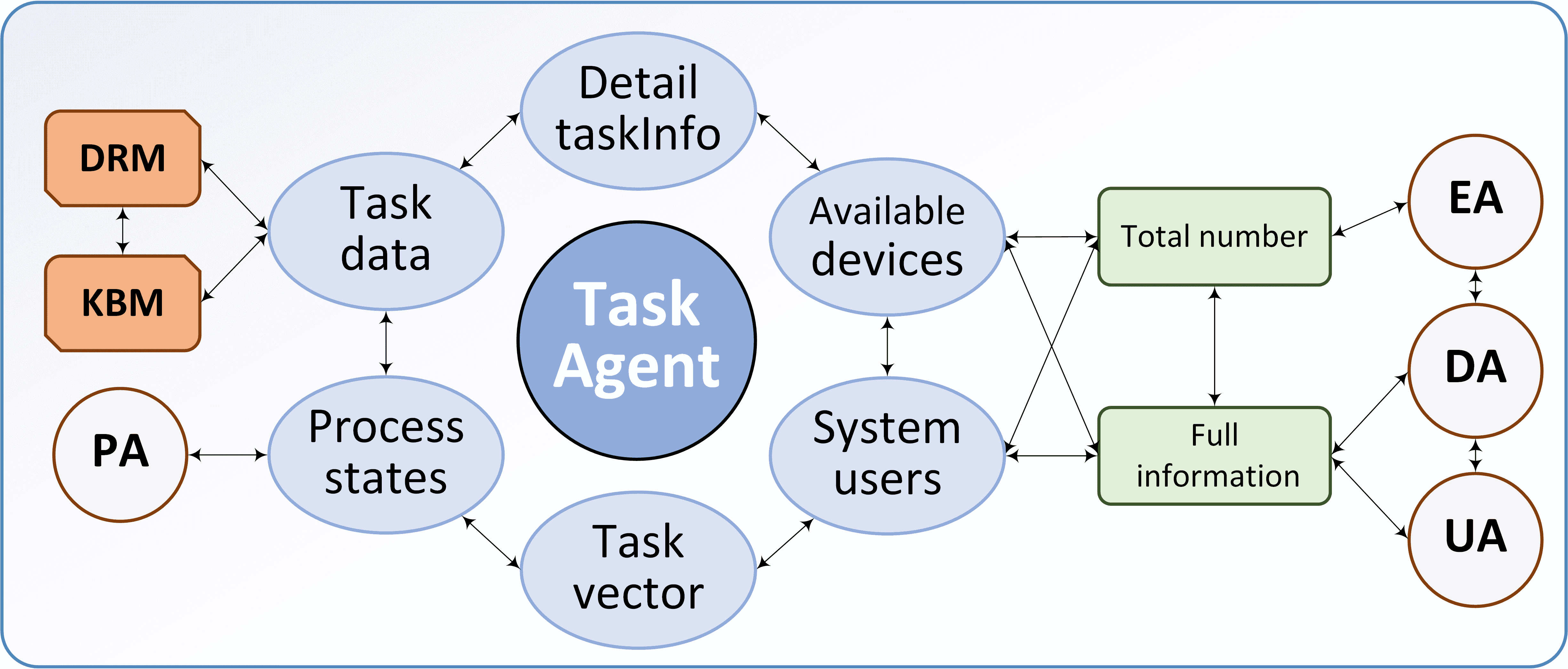}
\caption{Task resource graph T$_i$RG.}
\label{fig5}
\end{figure}

To complete the assignment process, we first need to control the resources that are global to the system, which can be obtained through the five agents. The resource graph construction process is as follows. Firstly, we need to detect the number of users and the total number of available devices in the \emph{CAP}. These can be obtained directly from the \emph{Environment-agent} structure. Secondly, sensing device status and user information in the current system are checked. These are separately stored in \emph{DA} and \emph{UA}. The task process status is available from \emph{PA}. From \emph{TA}, we can get detailed information about the current task. Lastly, based on analyzing and reasoning of these task-related information extracted from \emph{SRG}, the unique Task Resource Graph (\emph{TRG}) is constructed for each task, as shown in Fig. \ref{fig5}.

After generating \emph{TRG}, the system automatically assigns tasks to the appropriate participants. As shown in Fig. \ref{fig4} (c), scheduling and assignment mechanism includes three parts: strategy library, mapping model, and strategy management module. By analyzing and inferring the content of \emph{TRG}, the system can map the \emph{TID} to a specific strategy in library, thereby completing the process of strategy selection. Thus the conversion process \emph{P2} in Fig. \ref{fig4} is completed. The system then performs scheduling operations on devices or participants based on the selected strategy and assigns the task to appropriate executors. The strategy library stores commonly used or customized task assignment algorithmic functions, such as location-based, interest points based, game theory-based, and genetic algorithm-based algorithmic functions.

\begin{figure}[!tp]
\centering
\setlength{\abovecaptionskip}{0.cm}
\setlength{\belowcaptionskip}{-0.cm}
\includegraphics[width=3in]{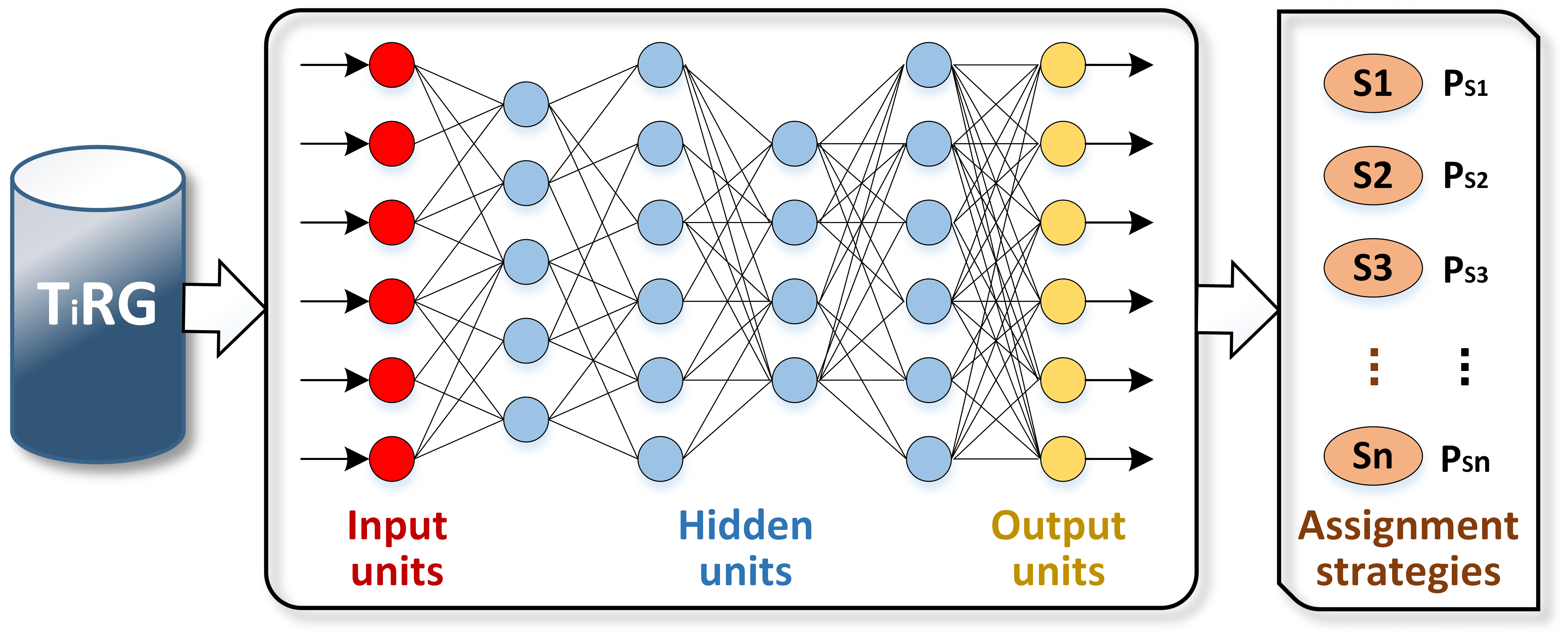}
\caption{ Scheduling and assignation strategy model.}
\label{fig6}
\end{figure}

The Scheduling and Assignation Strategy (\emph{SAS}) model is based on a deep neural network, as shown in Fig. \ref{fig6}, also known as the mapping model. It takes the probability of each strategy appropriate for the task as network outputs, the Resource Graph of $t_i$ (\emph{TiRG}) as inputs of network and establishes back propagation neural network model. In addition, the strategy management module is responsible for the management, revision and reconstruction of the strategy library and mapping model. Once the system is initialized, the policy library will automatically be incremented, modified, or removed periodically with the increasing of the number of processed tasks. The mapping model is also updated through online learning.

The top half of Fig. \ref{fig4} (c) shows the network topology for task assignment. Red circles represent different strategies in the library, blue circles are on behalf of tasks, and yellow circles stand for users or devices in the platform. As tasks enter the system, they are resolved in depth and mapped to the appropriate allocation strategies through a series of processes and eventually assigned to users or devices. Based on the combined effects of \emph{SAS} model, \emph{TRG} and the strategy library, the red, yellow, and blue circles are adaptively connected. There are many-to-one, one-to-one, and many-to-many relationships between tasks and strategies.


\section{Integrated Resource Management}

As a core framework of CrowdOS, Integrated Resource Management (\emph{IRM}) is a necessity for system stability and sustainability. Its specific contents include: users, smart terminals, physical computers, system environments, task processes, task data, and knowledge base. Generally, tasks needed to be handled are infinite but the computing resources are limited in system. To alleviate such a conflict and make sure the system resources are utilized rationally, we provide a uniform resource management mechanism. \emph{RM}  can provide abstract and unified management to resources.

\subsection{Agents Management}

Only if device, human and environment are united to be a large-scale system by correct management, reliability management of the whole system could be realized. When the sensing terminal is connected to the server-end, it will be triggered by the signal, and then the terminal automatically delivers the current device status to the server, such as device type, remaining power, location, real-time usage, as well as storage occupancy rate. The information can be captured and stored in device-agent, which assist in the realization of system functions, such as resource maximization and task scheduling, thereby helping the system to manage device resources in a fine-grained and organized manner.

System depicts the portrait of user through the \emph{User-agent}. Users mainly include two types, task participants and publishers. Both types of users rely on devices to interact with the system, for example, publishers release tasks through Human-Machine Interface (\emph{HMI}) of \emph{CAP} on smartphones. \emph{User-agent} not only stores common features such as user name, age, and related tasks, but also generates personalized information such as user credit rating, user preferences, and interest points.

Environmental resource is a collection of hardware and software resources of the server. It records the server architecture and processing power, such as centralized, distributed or edged deployment architecture, CPU numbers, CPU utilization, memory usage, available disk space. These resources are stored in the \emph{Environment-agent} and updated periodically to ensure that the system gets the latest data. \emph{EA} has an alarm function, which will predict according to the current system status and the increase and decrease of task size. If CPU utilization or storage usage reaches the rated threshold, the alarm will go off. The system automatically assigns or migrates task data to a distributed or edge server certain conditions are satisfied.

\subsection{Task Process Scheduling and Management}

Process-agent (\emph{PA}) is similar to the \emph{Process Control Block} in the operating system, and is a collection of phase state for the current task. The system assigns each task a unique process identifier (\emph{TPID}), which accompany the entire life cycle of the task. \emph{TPID} is stored in both \emph{TA} and \emph{PA}. \emph{PA} contains a wealth of information. For instance, \emph{TPID} is the unique identifier of the process. \emph{Process-state} describes the state of the current task in the system, there are seven switchable states. Process-strategy represents the process scheduling policy, such as \emph{FIFS}, \emph{RB}. \emph{Process-prio} means the process priority, from 0-15, which is in descending order, and numerical value 0 is defined as the highest priority. There are also other relevant information.

\begin{figure}[!tp]
\centering
\setlength{\abovecaptionskip}{0.cm}
\setlength{\belowcaptionskip}{-0.cm}
\includegraphics[width=3.5in]{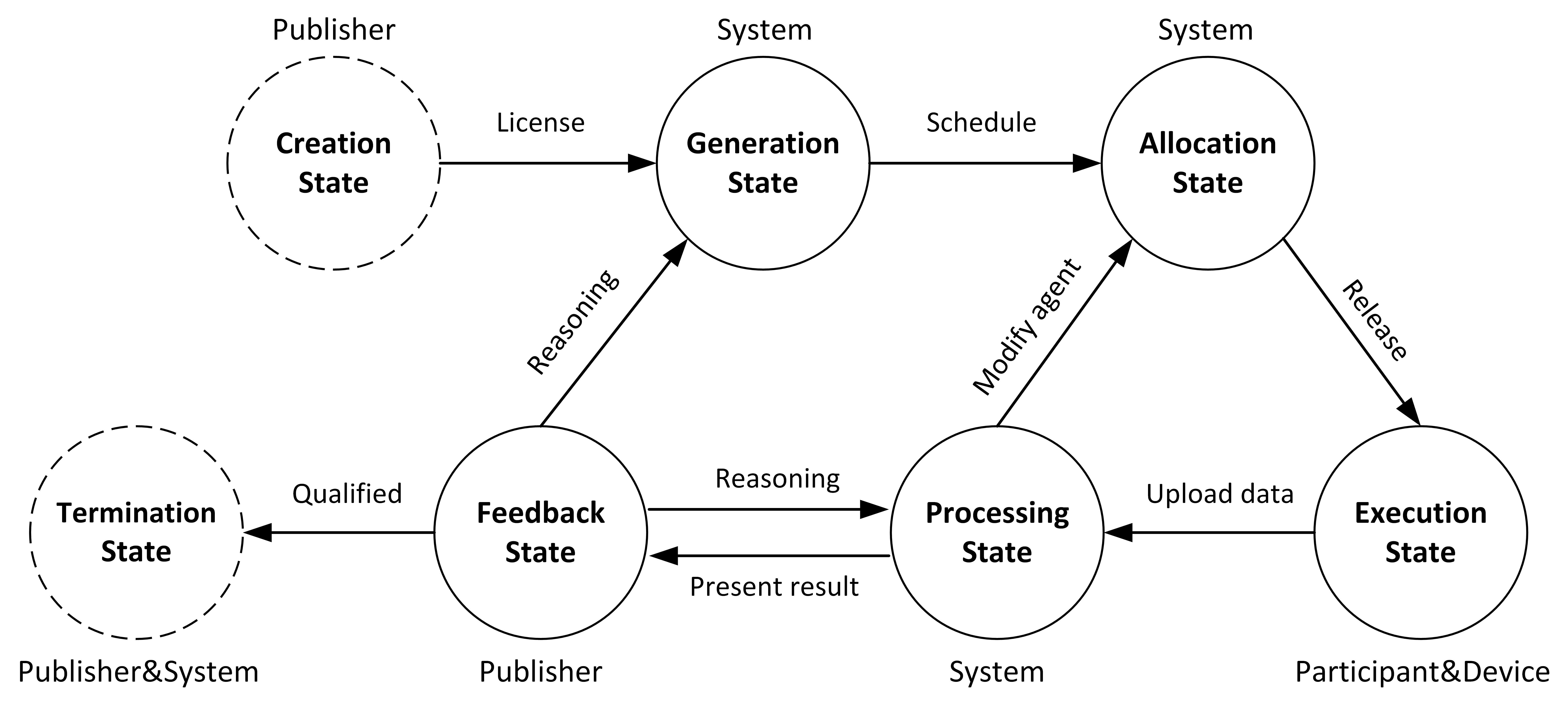}
\caption{The transition of task process state.}
\label{fig7}
\end{figure}

Task process has several states. In addition to the creation and termination states, there are also five states: generation, allocation, execution, processing, and feedback state. Fig. \ref{fig7} shows the transitions of these states.  Normally, a task goes from the creation state to the termination state clockwise. When the quality of task result is not qualified, the task will temporarily stay in the feedback state. After analyzing deep layer and surface causes, the task process may transition to the generation state, allocation state or processing state, and then continue execution. In Fig. \ref{fig7}, the text above the circle represent the execution entity of state: publisher, system, or participant. The text on the arrows indicate the action required to go from one state to another.

There are various integrated task process scheduling algorithms. 1) First-Come-First-Served (\emph{FCFS}), which prioritizes tasks that first enter the system, providing resources and services. 2) Round-robin, which generates a task interrupt at a periodic interval, and places the currently running process in the task-ready queue, then selects the next ready process based on \emph{FCFS}. 3) Task priority, tasks are processed in priority order, and the same level of tasks are scheduled in FIFO order. 4) \emph{HRRN}, the highest response ratio is prioritized, $R=(w+s)/s$, where $R$ represents the response ratio, $w$ means waiting time, $s$ is on half of the time expected to be served. 5) Feedback priority. For tasks in the feedback state, it would raise the priority level from $n$ to $n+2$. The selection of algorithm is based on the algorithm-flag bit in \emph{PA}.

\subsection{Heterogeneous Multimodal Data Management}

There are two types of data resources. The first is the intrinsic data carried by tasks, which we call Raw Data (\emph{RD}). The second is the New Uploading Data (\emph{NUD}) from the participants during the execution of tasks. The \emph{RD} includes text, image or voice that describe tasks, or data sets that need to be tagged by participant. These data will be presented to participants as tasks are released. The \emph{NUD} includes various types of sensory data uploaded by participants, such as text descriptions, sensor data, statistical charts, design documents and tagged data. The data structure in system is complex and diverse, including structured data, semi-structured data and a large amount of unstructured data.

Data management is a systematic project. Firstly, collecting and storing data is the first step, then it is necessary to build a retrieval method for unstructured data based on crowd intelligence. With these data under our belts, we can analyze them and try to build an indexing engine for the unstructured data. Secondly, data cube technology will be used to manage and store task data, we also construct a multi-feature cube for crowd data. In order to facilitate the search of data, we try to implement a complex query method based on multi-dimensional features, which helps to analyze the task data in a targeted manner. Thirdly, multidimensional data mining in cube space can combine the data from different tasks to provide support for discovering knowledge from massive tasks. It helps to discover knowledge in large-scale and semi-structured data sets, thereby leveraging data resources effectively. Last but not least, we provide a more flexible approach to retrieval and management for fine-grained analysis of massive amounts of data resources. With the explosive increase of task data, completed tasks and intermediate data will be periodically cleaned up automatically, and the useful or analyzed data will be transferred to the \emph{KB} for management.

\subsection{Knowledge Base Management}

Several major functions are required for knowledge base management, including knowledge operation and control, knowledge representation model, and knowledge search.

The knowledge of OS is divided into two categories: Existing Knowledge (\emph{EK}) and New Knowledge (\emph{NK}). \emph{NK} is extracted from tasks or data, which help to improve system mechanisms or update models. However, the information useful to users or third parties is not included in the scope of knowledge here. \emph{EK} includes expert strategies, decision rules or models defined in advance. For example, task allocation strategies already in library or reasoning trees. \emph{NK}, such as improved methods, patterns, addition rules, or updated network models, is often an extension of \emph{EK}. \emph{NK} mining means identifying potentially useful and interpretable information from \emph{EK} or data sets. Instead of knowledge explosion, the \emph{KB} size and complexity is limited to a controllable range.

Production rule, object-orientation, and frame are three main methods to represent knowledge in the OS. The method adopts unified algorithm numbering rule and knowledge expression based on produced rules to build task assignment strategy library and operation rule in \emph{KB}. Five-agents mechanism is based on object-oriented knowledge representation. Framework knowledge representation can store all the knowledge of an object to form a complex data structure. It plays an important role in achieving inexact reasoning of the \emph{KB}, such as the task resource graph, which is of great significance to the selection of assignment strategy.

Our knowledge search strategy is different from the common knowledge search engine, which is targeted for the crowdsourcing system and tasks. One of the important purposes of building a \emph{KB} is to effectively solve complex problems. The process of problem solving is essentially the knowledge matching and searching. \emph{KB} summarizes and stores knowledge based on its type, form, or level. Knowledge is not stored centrally in one management list or module, rather, it is distributed in various spaces. \emph{KB} mainly records the knowledge address, the inherent relationship between knowledge, and builds a network based on these.


\section{Task Result Optimization Framework}

For quality assessment and optimization of task results, we propose multiple effective mechanisms and establish a unified framework that mimics the process of human thinking. We consider two major issues in results quality: the number of results is sparse, and the error rate exceeds the standard. The notations in this part are listed in Table \ref{table2}.

\vspace{-0.25cm}
\begin{table}[!thbp]\footnotesize
\newcommand{\tabincell}[2]{\begin{tabular}{@{}#1@{}}#2\end{tabular}}
\caption{The Frequently Used Notations}
\label{table2}
\centering
\vspace{-0.25cm}
\begin{tabular}{p{1.1cm}p{5.1cm}}
\toprule
\multicolumn{1}{l}{\textbf{Symbol}} & {\textbf{Explanation} }\\
\midrule
{\tabincell{l}{DFHMI}}&{\tabincell{l}{Deep Feedback framework based on \\Human-Machine Interaction}} \\
{\tabincell{l}{QAM}} & {\tabincell{l}{Quality Assessment Method}}\\
{\tabincell{l}{PRQP}} & {\tabincell{l}{Possible Reasons of Quality Problems}}\\
{\tabincell{l}{PCL}} & {\tabincell{l}{Problem Causes Library}}\\
{\tabincell{l}{RN}} & {\tabincell{l}{Number of each Reason in PCL}}\\
{\tabincell{l}{SDIM}} & {\tabincell{l}{Shallow and Deep Inference Mechanism}}\\
{\tabincell{l}{DSP}} & {\tabincell{l}{Data Sparse Problem}}\\
{\tabincell{l}{LQR}} & {\tabincell{l}{Low Quality Result}}\\
{\tabincell{l}{RDT}} & {\tabincell{l}{Reason Decision Tree}}\\
{\tabincell{l}{SCOL}} & {\tabincell{l}{System Correction Operation Library}}\\
{\tabincell{l}{ON}} & {\tabincell{l}{Number of each Operation in SCOL}}\\
{\tabincell{l}{RSMT}} & {\tabincell{l}{RDT-SCOL Mapping Table}}\\
{\tabincell{l}{RNNM}} & {\tabincell{l}{Reason Neural Network Model for SDIM}}\\
\bottomrule
\end{tabular}
\end{table}
\vspace{-0.1cm}

\subsection{DFHMI Overview}

Currently, during the final results display phase, \emph{CAP} is mainly responsible for information aggregation but does not provide functional mechanism for recalibrating results. Due to the diversity and divergence of tasks and results, there was no uniform framework for evaluating and optimizing results quality. Therefore, we designed \emph{DFHMI}. The implementation of \emph{DFHMI} relies on Agents mechanism, assisting users to interact with system at a deeper level. In addition, it can not only update and improve internal decision models, but also enhance expandable capability of system in dealing with complicated problems.

\begin{figure}[thp]
\centering
\setlength{\abovecaptionskip}{0.cm}
\setlength{\belowcaptionskip}{-0.cm}
\includegraphics[width=3.5in]{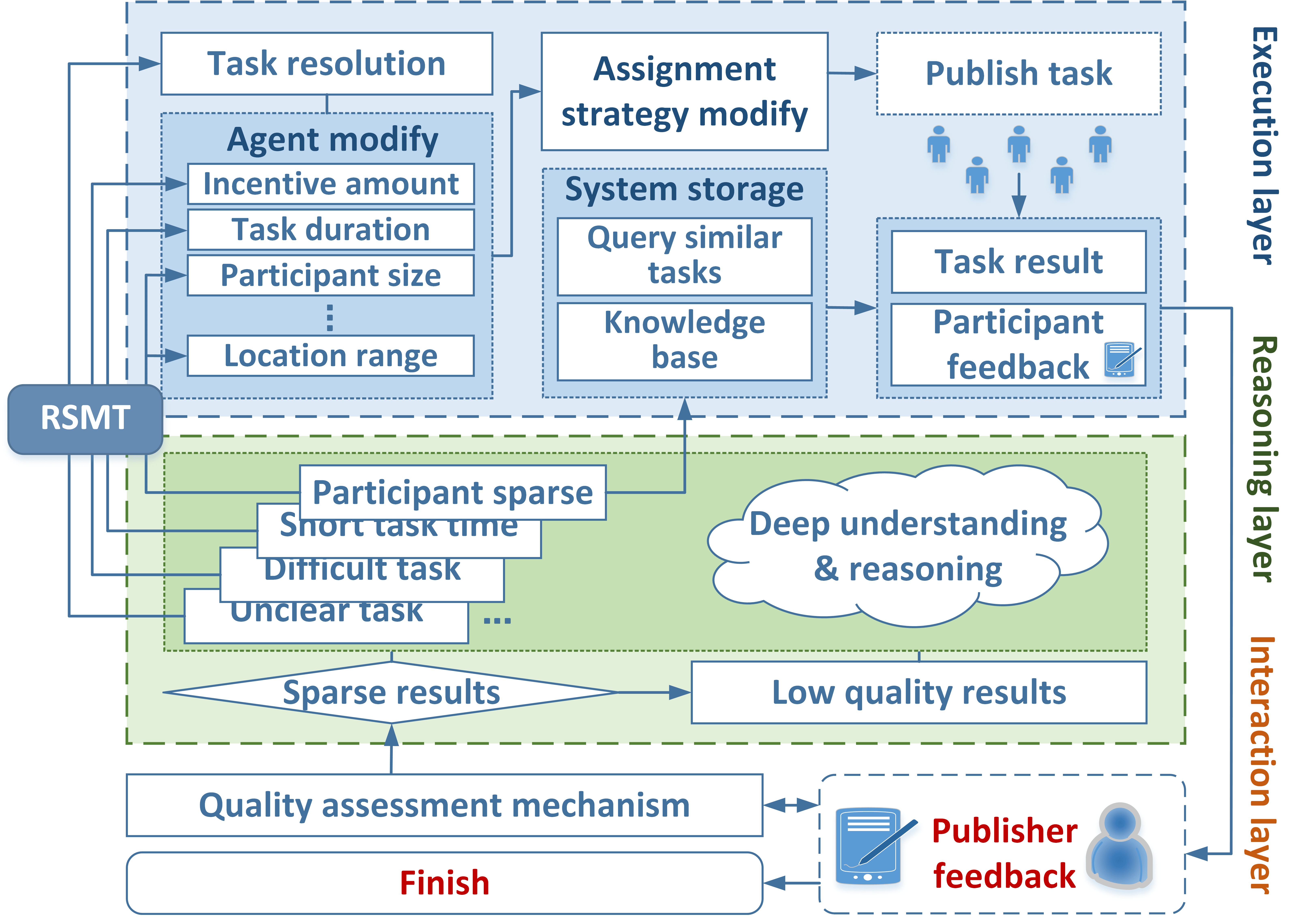}
\caption{Deep feedback framework based on human-machine interaction.}
\label{fig8}
\end{figure}

 In Fig. \ref{fig8}, we can see that \emph{DFHMI} consists of Interaction, Reasoning and Execution layers. In Interaction layer, the publisher evaluates the quality of task results through quality assessment method and upload results by entering evaluation information or clicking related buttons on the application interface. If the assessment result is better than the pre-determined threshold, the task will be terminated. If not, enter into the next layer. \emph{QAM} will be covered in more depth in Section 6.2. In Reasoning layer, the system performs key information extraction and in-depth analysis operations of the feedback information to explore the Possible Reasons of Quality Problems (\emph{PRQP}). Based on the reasoning model established, all obvious and deeper reasons are mapped to the Problem Causes Library (\emph{PCL}). The system then compares \emph{PRQP} of each task with \emph{PCL}. \emph{RSMT} is the bridge between the reasoning and execution layers. The third layer is Execution layer, where every problem in \emph{PCL} corresponds to internal operations. We has already defined most of the operations initially. Part of these operations are implemented by modifying certain values in agents, there are also other types of operations such as switching process. After correction operations are completed, the task will enter into the new process phase. Finally, new results will be fed back to the publisher again, waiting for the interaction layer to re-evaluate them through \emph{QAM}. The entire optimization process forms a half loop and the task breaks out of the loop when results quality is qualified.

\subsection{Quality Assessment Mechanism}

We consider five principles for the quality assessment of task results and apply a quantifiabler approach to assist in evaluating. There are two levels of assessment, the overall or the independent, which means that the assessment is for the results submitted by all participants or each participant independently. Here we assume an holistic evaluation will be done for task $t_i$ results. We define the quality assessment function $Q_{t_i}$ according to the following principles.

\begin{itemize}
  \item Content relevance, $\xi_{t_i}^1$. Determine how relevant the result is to task $t_i$.
  \item Format correctness, $\xi_{t_i}^2$. Determine whether the format of uploaded data meets $t_i$ requirements.
  \item Non-redundancy, $\xi_{t_i}^3$. Determine if the participant has uploaded redundant data or repeated uploads.
  \item Completeness, $\xi_{t_i}^4$. Determine if the result contains all of the specified content.
  \item Submission speed, $\xi_{t_i}^5$. Determine how fast the results are submitted.
\end{itemize}

Each $\xi_{t_i}^k$, $k\in(1,\dots,5)$ represents a value between 0 and 1, where $\xi_{t_i}^k={S_{t_i}^k}/{\Phi_{t_i}^k}$. The numerator $S_{t_i}^k$ is the score given by the publisher according to the above principles, and the denominator $\Phi_{t_i}^k$ is the upper limit of score value. The quality assessment function is as follows:

\textbf{Definition 4 (Result Quality Assessment Function).}

\begin{small}
\begin{equation} \label{eq2}
Q_{t_i}=\frac{\sum_{k=1}^5 {{\xi_{t_i}^k}\cdot{\omega_{t_i}^k}}}{1+E_{t_i}}
\end{equation}
\end{small}

$\omega_{t_i}^k$ is the weight of the corresponding $\xi_{t_i}^k$, where $\omega_{t_i}^1+\cdots+\omega_{t_i}^5=1$ in Eq. (\ref{eq2}). The distribution ratio of $\omega_{t_i}$ is obtained according to the analysis of $t_i$, such as types, range, or other features. The numerator of Eq. (\ref{eq2}) is the total score of results. The value $1$ in denominator comes from $\sum_{k=1}^5 {{(\xi\to \Phi)_{t_i}^k}/\Phi_{t_i}^k\cdot{\omega_k}}$. $E_{t_i}$ is estimating entropy of $t_i$.

\textbf{Definition 5 (Estimating Entropy).} Estimating entropy $E_{t_i}$ represents the purity of the crowdsourcing task results. The larger the $E_{t_i}$, the greater the amount of information required to complete the assessment, which increase the uncertainty in assessment process. The formula of estimating entropy is shown in Eq. (\ref{eq3}).

\begin{small}
\begin{equation} \label{eq3}
E_{t_i}=F(\frac{\Lambda_{t_i}}{\Upsilon_{t_i}})
\end{equation}
\end{small}

$\Lambda_{t_i}$ is the amount of additional information required to evaluate $t_i$, with the information increase, the value of $\Lambda_{t_i}$ increases. While $\Upsilon_{t_i}$ means the accuracy and ambiguity of the description of $t_i$. The higher the accuracy and the smaller the ambiguity, the higher the $\Upsilon$. ${F(x)}_{normalize}=\frac{1}{1+{e}^{-\log x}}$.

Therefore, as the entropy rises, the reliability of result quality assessment reduced, further affecting and reducing $Q_{t_i}$ value in Eq. (\ref{eq2}). When the value of $Q_{t_i}$ is greater than a previously set threshold $\delta$, that is, $Q_{t_i}\ge\delta$, the result optimization process is activated. $\delta$ can be fine adjusted according to practical circumstances. The higher the value, the better the result quality. Based on the principles and assessment function, we can not only evaluate the quality of results, but also explore the deep reasons and put forward the operational strategy to improve results.

There are three reasons for using the evaluation method based on human-machine synergy. First of all, each task in the system is independent. Due to the variety of data formats and fields presented by the results of different tasks, the system cannot accurately measure the error rate through a unified automatic evaluation method. Secondly, the results of crowdsourcing tasks are mostly composed of unstructured data. There is no general direct analysis method for heterogeneous unstructured data. Unless the system performs in-depth processing on each task to extract useful structured information, no more accurate judgment can be made. Also, this operation consumes a lot of system resources. Therefore, the method of measuring the quality of the results by analyzing the data itself is not feasible. Thirdly, when the error rate is calculated by the above evaluation principle, the score given by the task publisher needs to be taken as the main reference.

\subsection{Shallow-Deep Inference Mechanism}

In order to identify the reasons for quality problems and explore solutions, we design the \emph{SDIM}. The main process is as follows. The system first establishes a Reason Decision Tree (\emph{RDT}) for the $Q\leq\delta$ task. The shallow cause combined with the task characteristics can infer the underlying cause. We then design a System Correction Operation Library (\emph{SCOL}) to address the underlying reason. The mapping relationship between nodes of \emph{RDT} and \emph{SCOL} is stored in the system as an \emph{RDT-SCOL} Mapping Table (\emph{RSMT}).

There are two types of \emph{RDT}: global \emph{RDT} and task \emph{RDT}. We elaborate the global \emph{RDT} construction process first. All shallow and deep causes reside in \emph{PCL} and each cause has a corresponding Reason Number (\emph{RN}). Generally, deep causes are more elaborate than shallow ones, but there are no clearly distinct boundaries between them and sometimes they are interchangeable. The root node of global \emph{RDT} represents \emph{PCL} and the child nodes of root node consist of shallow causes. The new shallow reason can be added as the child node of the root node. Deep causes are based on shallow causes, and are the extension of the nodes in the longitudinal direction. With the scale of the \emph{PCL} increases, in order to improve accuracy and reduce retrieval times, \emph{RDT} will perform the branching or update operations periodically to ensure it is maintained within a certain scale. For nodes that have not been retrieved for a long time, they can be pruned. For reasons that are not searchable in the tree, we can add child nodes under the appropriate parent node through reasoning. Global \emph{RDT} is a pure decision tree model and can be constructed by calculating the conditional probability. However, task \emph{RDT} is an integrated model that combines global \emph{RDT} and neural networks.

The ultimate goal of \emph{SDIM} is to find the final solution and optimize the results for each task with quality issues. Shallow causes are easily understood or directly obtained from feedback. For task \emph{RDT}, underlying or deep causes are usually not directly accessible but obtained by Reason Neural Network Model (\emph{RNNM}). Inputs of \emph{RNNM} are combination of shallow causes and task-related features, while the outputs are deep causes. First, we find the essential cause of the task result problem and the corresponding \emph{RN} by building task \emph{RDT}. Second, we map \emph{RN} to \emph{ON} through \emph{RSMT}. Last, correction bit in the \emph{Task-agent} is filled by \emph{ON} and the correction operations are activated automatically. The following are specific instructions of \emph{SDIM} through Data Sparse and Low Quality Result problems.

\subsubsection{Compensation Strategy for Data Sparsity}

\begin{table}[!htp]\footnotesize
\newcommand{\tabincell}[2]{\begin{tabular}{@{}#1@{}}#2\end{tabular}}
\renewcommand{\arraystretch}{1.3}
\caption{RDT-SCOL mapping table for DSP}
\label{table3}
\centering
\vspace{-0.25cm}
\begin{tabular}{p{0.4cm}p{1.7cm}p{0.4cm}p{2.2cm}p{2.1cm}}
\toprule
\tabincell{l}{\textbf{RN}} & \tabincell{l}{\textbf{Reason-phrase}} & \tabincell{l}{\textbf{ON}}&\tabincell{l}{\textbf{Revision-phrase}}& \tabincell{l}{\textbf{Operations}}\\
\midrule
\tabincell{l}{\textbf{0x01}} & \tabincell{l}{\ unclear task\\\ description } & \tabincell{l}{\textbf{No.1}} & \tabincell{l}{\ re-decompose and \\\ generate tasks} & \tabincell{l}{PS-$>$\\Generated state}\\
\tabincell{l}{\textbf{0x02}} & \tabincell{l}{\ insufficient\\\ participants} & \tabincell{l}{\textbf{No.2}} & \tabincell{l}{\ increase number \\\ of participants } & \tabincell{l}{M-$>$\\Task-agent.range}\\
\tabincell{l}{\textbf{0x03}} & \tabincell{l}{\ difficult task} & \tabincell{l}{\textbf{No.3}} & \tabincell{l}{\ increase \\\ incentives} & \tabincell{l}{M-$>$\\Task-agent.reward}\\
\bottomrule
\end{tabular}
\end{table}

\begin{table*}[thb]\footnotesize
\newcommand{\tabincell}[2]{\begin{tabular}{@{}#1@{}}#2\end{tabular}}
\renewcommand{\arraystretch}{1.3}
\caption{RDT-SCOL Mapping Table for LQR}
\label{table4}
\centering
\vspace{-0.25cm}
\begin{tabular}{p{0.9cm}p{3.8cm}p{0.9cm}p{5.8cm}p{3.4cm}}
\toprule
\textbf{RN} & \textbf{Reason-phrase} &\textbf{ON}& \textbf{Revision-phrase}& \textbf{Operations}\\
\midrule
\textbf{0x1001}&\tabincell{l}{no relevant between result\\ and task} &\tabincell{l}{\textbf{No.31}\\\textbf{No.4}}& \tabincell{l}{Filter out high-credit users from participants\\ Republish task}&\tabincell{l}{M-$>$Task-agent.u-credit\\PS-$>$Assignment state}\\
\textbf{0x1002} &\tabincell{l}{result format is not uniform} & \tabincell{l}{\textbf{No.32}\\\textbf{No.5}} &\tabincell{l}{modify the result format requirement \\request user to submit results again} & \tabincell{l}{M-$>$Task-agent.format\\M-$>$Task-agent.submit-state}\\
\textbf{0x1003} &\tabincell{l}{redundant results or repeated\\ data uploads} & \tabincell{l}{\textbf{No.33}\\\textbf{No.31}} & \tabincell{l}{warn users and lower their credit rating \\ Filter out high-credit users from participants} & \tabincell{l}{M-$>$User-agent.credit\\M-$>$Task-agent.u-credit}\\
\bottomrule
\end{tabular}
\end{table*}

First of all, we need to reason out the essential cause (i.e. special deep \emph{RN} for the task) for $t_i$ from the shallow cause through task \emph{RDT}. The causes of \emph{DSP} may include insufficient number of participants, too short task execution time, difficult task, unclear task descriptions, each of which corresponds to a \emph{RN}, as shown in Table \ref{table3}. Reason-phrase was obtained from the feedback information through natural language processing. For a shallow node with insufficient number of participants, the corresponding deep cause or child node could be less task incentives, too small a task release range, etc. Combined with task features, we can infer the most likely underlying cause through \emph{RNNM}.

Secondly, we need to find the correction method for $t_i$ (i.e., special \emph{ON} for $t_i$), which can be achieved by \emph{RSMT}. Table \ref{table3} lists several causes in \emph{RN} and operations in \emph{ON} for \emph{Data Sparse Problem}. \emph{RN} and \emph{ON} are freely mapped by \emph{RSMT}, and the correspondence between them includes one-to-one, many-to-one, and many-to-many. The list of \emph{RN} and \emph{ON} are updated regularly as new issues arise. We explain the symbolic meaning in Table \ref{table3}. For example, the fourth row and third column record the item of increasing incentive amount. The corresponding system operation is \emph{M-$>$Task-agent.reward}, \emph{M} means modify, and \emph{increase} means increasing the value of reward in \emph{Task-agent}. \emph{PS} represents the state of the current task process, while \emph{PS-$>$generated state} means that the process is switched to the task generation state.

Thirdly, we describe how \emph{SDIM} helps the system make right decisions in \emph{DSP}. If the direct reason is insufficient participants and there is no deep reasoning layer, then the system is likely to give a correction method as raising the incentive to attract more participants. With \emph{SDIM}, the system can combine shallow reasons with task information to make a comprehensive judgment. Suppose we extract the similar words or phrases as location, scope, execution location or remoteness from the feedback information, it is likely that the number of participants is sparse because the task execution scope is small. Here the right correction method is to expand the scope of the execution target and then republish the task. In this example, if the underlying reasons are correct, then the correction strategy given by the shallow reason may lead us in the wrong direction.

Last but not least, there are multiple types of revision methods in the system, which are in different levels or functional modules of the OS kernel. For the result sparsisty problem, the above compensation strategy is effective. The second method is to find the completed task in the system repository, and initiate a data call request for the task whose similarity with the current task reaches the specific threshold, and use the result as the compensation data. The third way is to search for task results on the network, integrate existing data through machine learning or rule methods, and organize them into task results.

\subsubsection{Remediation Strategy for Low Quality Result}

In this section, we mainly address the problem of \emph{LQR}. Suppose that a task publisher has received sufficient task results, but the data quality is not qualified. Crowdsourcing tasks involve many uncertain factors in the implementation process, but the causes of problems are often difficult to predict. If the cause is not inferred in advance, it is difficult to accurately give the correction and optimization method. Therefore, we need to analyze the uncertain factors and find out reasons that are most likely to cause \emph{LQR} problem. For example, the reasons may be that task description has semantic differences, the submitted data format does not meet the requirements, or the participants falsify data results.

There are three important operations that need to be completed to establish the correction mechanism for $t_j$. The first is to complete the contents in \emph{PCL} and find out the specific \emph{RN} of $t_j$ by the task \emph{RDT}. The second is to correspond \emph{RN} to \emph{ON} in \emph{SCOL} through \emph{RSMT}. The initial design of \emph{PCL}, \emph{SCOL} and \emph{RSMT} are all based on prior knowledge. However, they are scalable and can be optimized and modified as the amount of task increases. Third, activate the system operations corresponding to the obtained operation numbers for $t_j$.

As shown in Table \ref{table4}, $M$ represents the system correction operation and is followed by the content to be corrected. Each content meaning can be viewed by the explanation of corresponding bits in \emph{agent} classes. \emph{PS} indicates that the current task process needs to be switched. For the specific meaning of the switching status, refer to Section 5.2.

Once \emph{DFHMI} is enabled, the system will continuously adjust models and parameters based on the combination of initial state and current information. The models and strategies in \emph{DFHMI} will be continuously updated and optimized. Gradually, the accuracy to solve problems is improved and the decision time is shortened. More importantly, the system supplements new knowledge through self-learning and human-machine collaboration, and then migrates the learned knowledge to new tasks ceaselessly.

\section{Key Components and Interfaces}

\begin{table}[!thbp]\footnotesize
\newcommand{\tabincell}[2]{\begin{tabular}{@{}#1@{}}#2\end{tabular}}
\caption{Extensible Components, Libraries, and Models}
\label{table5}
\centering
\vspace{-0.25cm}
\begin{tabular}{p{1.5cm}p{1.1cm}p{5.0cm}}
\toprule
\multicolumn{1}{l}{\textbf{Attribute}} & \textbf{Symbol} & \textbf{Content}\\
\midrule
{\tabincell{l}{Functional\\components}}&{\tabincell{l}{SPSM\\IMM\\CTDM\\CEM\\QAM}}&{\tabincell{l}{Strong Privacy Security Module\\Incentive Mechanism Module\\Complex Task Decomposition Module\\Credit Evaluation Module\\Quality Assessment module/method}} \\
\midrule
{\tabincell{l}{Libraries\\and\\models}} & {\tabincell{l}{TASL\\TRM\\PCL\\RDT\\SCOL\\RSMT\\RNNM}} & {\tabincell{l}{Task Assignation Strategy Library\\Task Resource Map\\Problem Causes Library\\Reason Decision Tree\\System Correction Operation Library\\RDT-SCOL Mapping Table\\Reason Neural Network Model for SDIM}}\\
\bottomrule
\end{tabular}
\end{table}

CrowdOS provides a rich set of Functional Components and Scalable Libraries. The completed part can be found in our project website \cite{crowdos.cn}. 

As shown in Table \ref{table5}, \emph{SPSM} can enhance the ability of system to protect user privacy, such as blockchain-based user location privacy protection mechanism. \emph{IMM} can mobilize the enthusiasm of users to participate in the task \cite{guo2017taskme}, \cite{nan2014cross}, and increase user engagement through a combination of cash incentives and virtual incentives. \emph{CTDM} can split tasks that require multiple steps into sub-tasks that can be executed in parallel. \emph{CEM} divides the credit rating for the user and improves the quality of the task result. \emph{QAM} is embedded in \emph{DFHMI} and can be updated separately without compiling the entire system kernel. Personalized and selectable  components not only enrich system functions, enhance user experience, but also help the system run accurately and efficiently.

These are internal system modules, libraries and models with a rich set of interfaces. \emph{TASL} contains a variety of basic assignment algorithms and special algorithms for special scenarios \cite{guo2016activecrowd}, \cite{liu2016taskme}, and new assignment algorithms can be added to the library according to the protocol. \emph{TRM} includes all the resources available in the system. In Section 6, we have fully demonstrated the \emph{RDT}, \emph{SCOL}, \emph{RSMT} and \emph{RNNM} which are the core of \emph{DFHMI}. We also provide interfaces through which the components and models can communicate with each other.

Besides rich internal interfaces, we also expose external callable interfaces (CrowdAPI), which are open to third-party application developers. By leveraging the services provided by CrowdOS, application developers only need to handle high-level business logic, thus simplifying their work significantly.

\begin{table}[!tbp]\footnotesize
\newcommand{\tabincell}[2]{\begin{tabular}{@{}#1@{}}#2\end{tabular}}
\caption{Application Execution and Development Environment}
\label{table6}
\centering
\vspace{-0.25cm}
\begin{tabular}{p{3.2cm}p{4.8cm}}
\toprule
\multicolumn{2}{l}{\textbf{Application Execution Environment}}\\
\midrule
{\tabincell{l}{Smartphone, CPU model}}&{\tabincell{l}{P30 ELE-AL00, Hisilicon Kirin 980}} \\
{\tabincell{l}{OS, Android version}} & {\tabincell{l}{Android 9.0, EMUI 9.1.0}} \\
{\tabincell{l}{Cores,GPU,RAM,Storage}} & {\tabincell{l}{8 core 2.6GHz, Mali-G76, 8 GB, 128 GB}} \\
{\tabincell{l}{Sensors}} & {\tabincell{l}{Gravity sensor, Ambient light sensor,\\ Gyroscope, Proximity sensor, GPS, etc.}} \\
{\tabincell{l}{Server OS}} & {\tabincell{l}{CentOS 6.9 64bit}} \\
\midrule
\multicolumn{2}{l}{\textbf{Development and Test Environment}}\\
\midrule
{\tabincell{l}{Front-end IDE}} & {\tabincell{l}{Android Studio 3.4.2}} \\
{\tabincell{l}{JRE, SDK}} & {\tabincell{l}{JRE 1.8.0, Android 9.0(Pie) API 29}} \\
{\tabincell{l}{Back-end IDE}} & {\tabincell{l}{IntelliJ IDEA Community Edition 2019}} \\
{\tabincell{l}{Postman, MySQL}} & {\tabincell{l}{Postman v7.2.2, MySQL 8.0.1}} \\
{\tabincell{l}{Navicat Premium}} & {\tabincell{l}{Navicat Premium 12.1}} \\
{\tabincell{l}{Maven, Spring}} & {\tabincell{l}{Apache-maven-3.6.1, SSM}} \\
{\tabincell{l}{SDK (ADB)}} & {\tabincell{l}{Android Debug Bridge version 1.0.41}} \\
\bottomrule
\end{tabular}
\end{table}

\section{Performance Evaluation}

We mainly evaluate CrowdOS  and the \emph{CAP} from four aspects: Correctness and Efficiency ($Ev_1$), Validity and Usability ($Ev_2$), Optimized Result Quality Assessments ($Ev_3$), Performance, Load and Stress Testing ($Ev_4$).

\subsection{Experiment Design}

We developed a \emph{CAP}, \emph{WeSense}. During the implementation of \emph{WeSense}, $Ev_1$ and $Ev_2$ are evaluated based on two development methods: $M_1$ (independent development), $M_2$ (CrowdAPI-based development). $Ev_3$ is an assessment of the core framework \emph{TRO}, while $Ev_4$ is the overall performance and stress test of \emph{WeSense} supported by CrowdOS. The environments are shown in Table \ref{table6}.

In $Ev_1$ and $Ev_2$, we compared and analyzed the time required to implement related functional modules under $M_1$ and $M_2$, and then split the functions and modules $F\{f_i\}$ to be tested into five parts ($F_{num}$=5). The empirical results and time consumption are described in later subsections.

${f_1}$: Task real-time publishing function.

${f_2}$: Task assignment algorithm module.

${f_3}$: Privacy protection module.

${f_4}$: Crowdsourced data collection and upload function.

${f_5}$: Result quality optimization function.

We set up the experimental environment and installed related software in advance. We hired nine volunteers (${\Lambda_{num}}$=9) who are familiar with the Java programming language and gave them two weeks to develop an app. We first spent 25 minutes ($time_{c1}$) to  introduce the functions of CrowdOS and all the APIs. We put nine volunteers in group A first, and then switch them to group B later, that is, each volunteer served as both $G_A$ and $G_B$ member at different times, $G_{A\_num}$=$G_{B\_num}$=9. 
$G_A$ member use $M_1$, while $G_B$ use $M_2$. Every volunteer participated in all tests, the total number of tests is $(G_{A\_num}+G_{B\_num})*F_{num}$=90.

The four evaluation metrics are compared as follows.
\begin{itemize}
\item $Ev_1$. We measure it by comparing the time consumption to complete $F\{f_i\}$ and the integrity and correctness of $F\{f_i\}$ developed through $M_1$ and $M_2$.

\item $Ev_2$. We compare the realization effect and time consumption of $F\{f_i\}$ performed by $G_A$ and $G_B$.

\item $Ev_3$. we compare the results through \emph{TRO} framework and other methods, the difference between the optimization effect and time consumption.

\item $Ev_4$. It is an overall system test with multiple quantitative factors.

\end{itemize}

\subsection{Correctness and Efficiency Assessment}

\begin{figure}[thbp]
\centering
\setlength{\abovecaptionskip}{0.cm}
\setlength{\belowcaptionskip}{-0.cm}
\subfigure[ ]{
\begin{minipage}[t]{0.25\linewidth}
  \includegraphics[width=\textwidth]{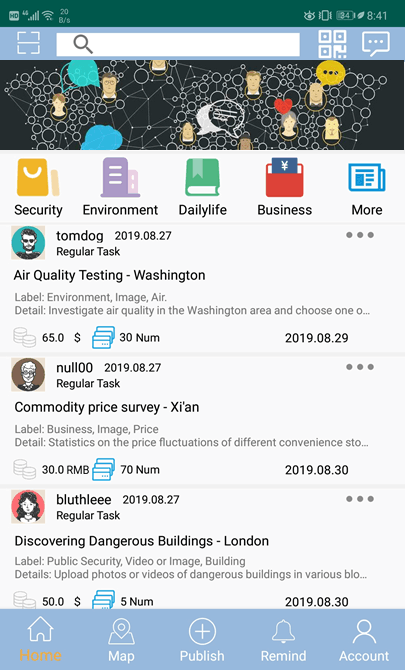}
\end{minipage}
}
\subfigure[ ]{
\begin{minipage}[t]{0.25\linewidth}
  \includegraphics[width=\textwidth]{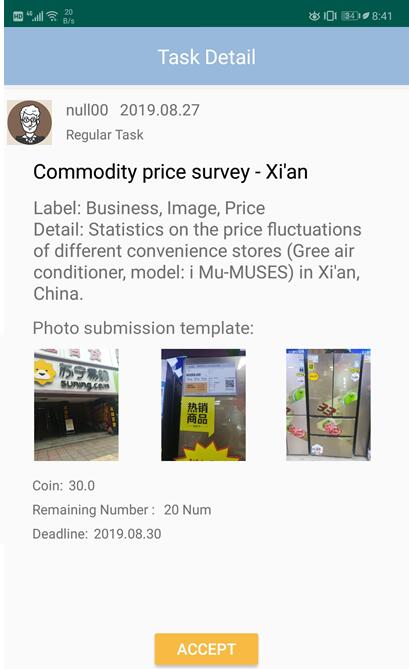}
\end{minipage}
}
\subfigure[ ]{
\begin{minipage}[t]{0.25\linewidth}
  \includegraphics[width=\textwidth]{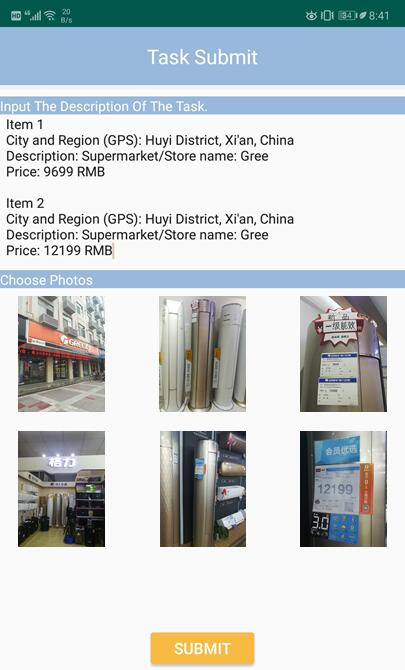}
\end{minipage}
}
\centering
\caption{\emph{WeSense} interfaces. (a) home page: crowd task display and search; (b) task detailed page: clicked to view the task of interest; (c) task submission page: submit the completed result data.}
\label{figx1}
\end{figure}

First, we removed the tests with a correct rate higher than the threshold, $C_{rate}>\delta_{c}$. According to the actual inspection and analysis, 90 tests have  passed the correctness screening. Fig. \ref{figx1} shows some of the user interfaces.

\begin{figure}[htbp]
\centering
\setlength{\abovecaptionskip}{0.cm}
\setlength{\belowcaptionskip}{-0.cm}

\subfigure[ ]{
\begin{minipage}[t]{0.49\linewidth}
  \centering
  \includegraphics[width=\textwidth]{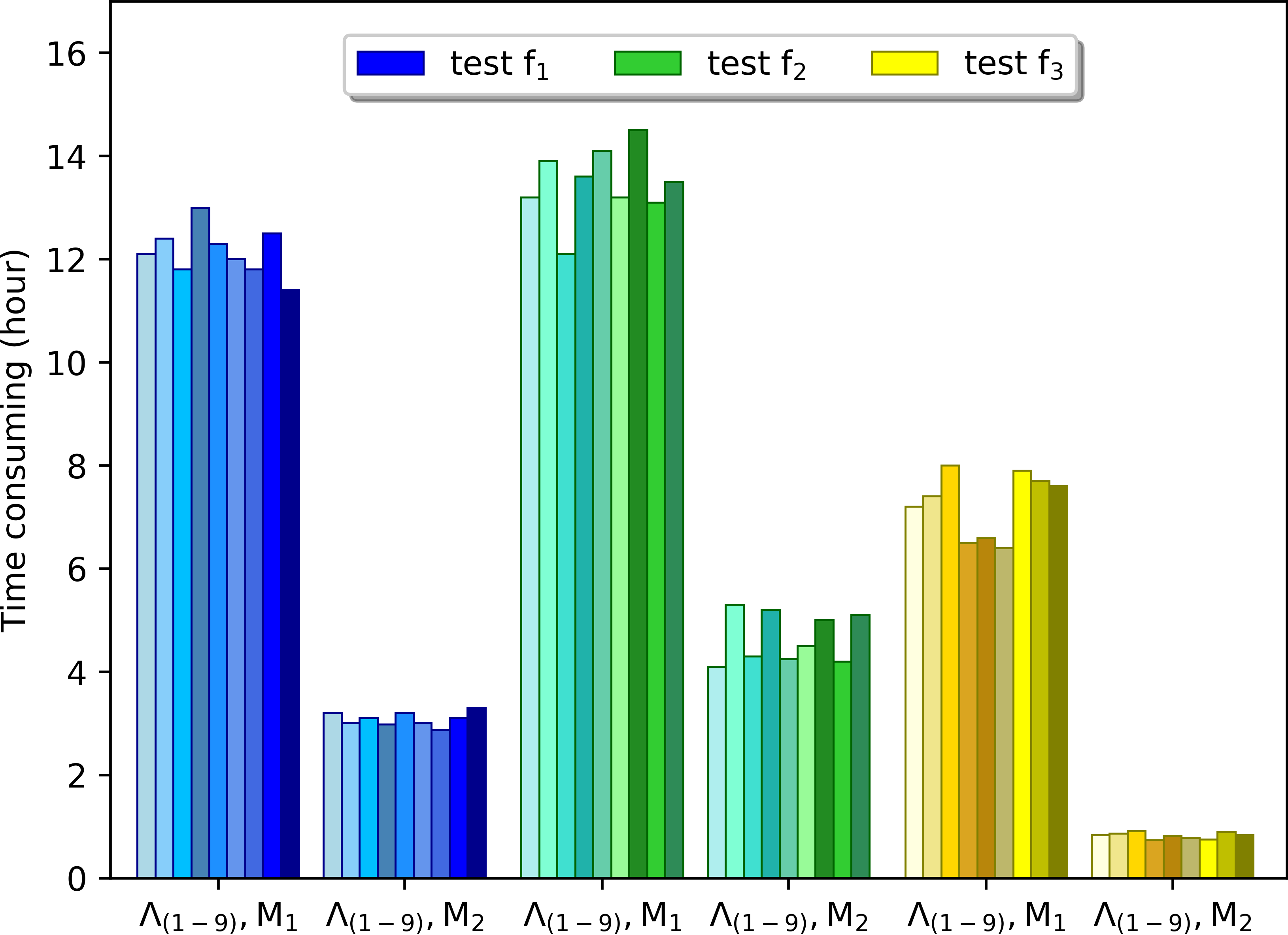}
\end{minipage}
}%
\subfigure[ ]{
\begin{minipage}[t]{0.49\linewidth}
  \includegraphics[width=\textwidth]{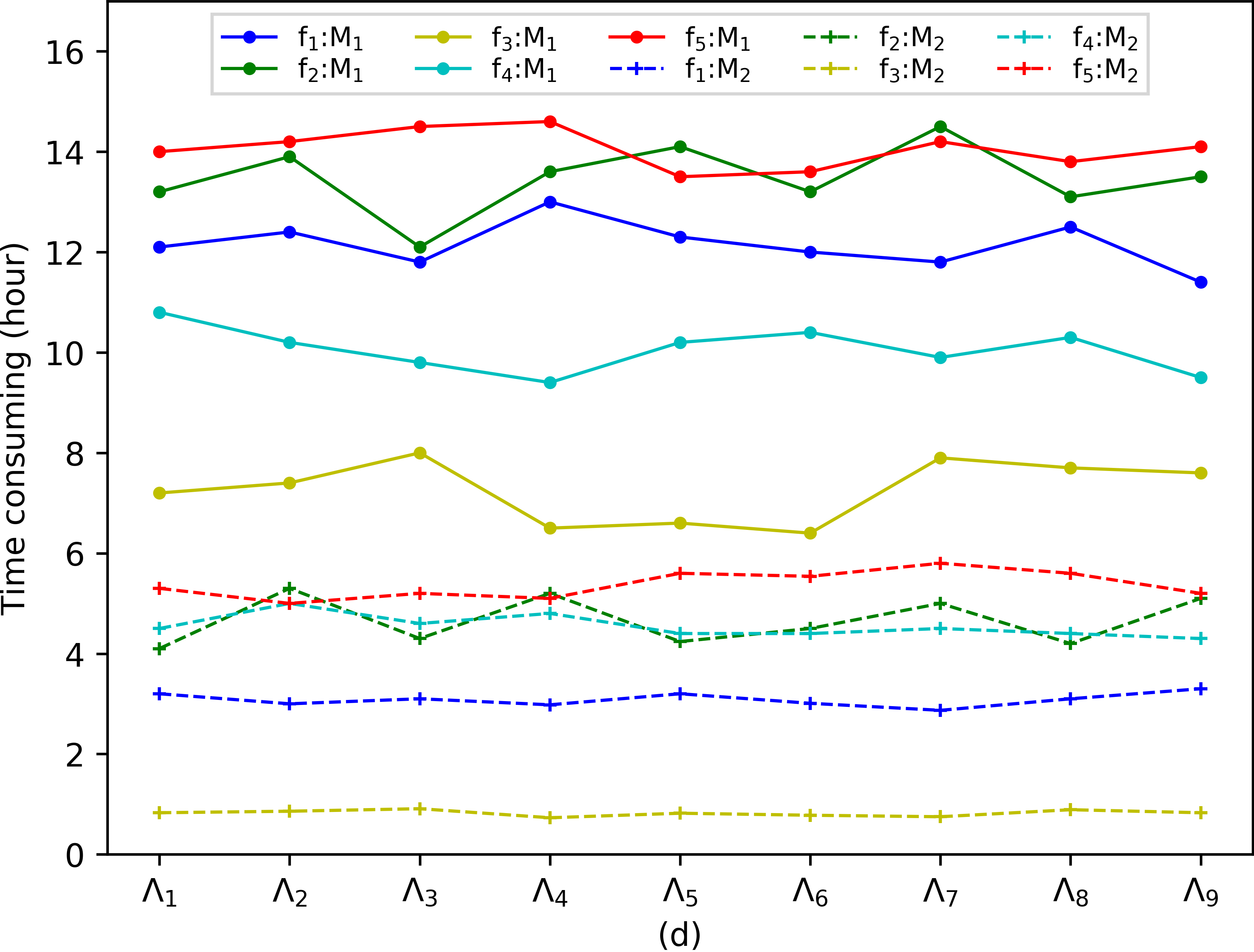}
\end{minipage}
}
\centering
\caption{Efficiency assessment. (a) time consumption of ${f_{1-3}}$; (b) overall evaluation $F$, $\Lambda$ - volunteer, $f_1:M_1$ - test f1 by development mode 1.}
\label{figx2}
\end{figure}

Analysis and comparison of the development cycle is shown in Fig. \ref{figx2}. Fig. \ref{figx2}(a) shows the time spent on completing the ${f_{1-3}}$ tests by $G_A$ and $G_B$. Fig. \ref{figx2}(b) shows the time consumption comparison of all tests based on $M_1$ and $M_2$ .

As shown in Fig. \ref{figx2} (b), the average development time of $F\{f_i\}$ is reduced from the original $DT_{G_A}(tc_{f_i})=\{12.1, 13.5, 7.3, 10.1, 14.1\}$ to $DT_{G_B}(tc_{f_i})=\{3.1, 4.7, 0.8, 4.5, 5.4\}$ hours, according to Eq. (\ref{eqx1}).

\begin{small}
\begin{equation} \label{eqx1}
DT{(tc_{f_i})}=\sum_{k\in \Lambda}{tc_{f_i}^{\lambda_k}}/\Lambda_{num}
\end{equation}
\end{small}

Therefore, the overall Development Efficiency ($DE$) of $f_1\to{f_5}$ increased by 310$\%$, according to Eq. (\ref{eqx2}):

\begin{small}
\begin{equation} \label{eqx2}
DE_{overall}={\sum_{i=1}^{F_{num}}\frac{{{DT_{G_A}{(tc_{f_i})}}-DT_{G_B}{(tc_{f_i})}}}{DT_{G_B}{(tc_{f_i})}}}/{F_{num}}
\end{equation}
\end{small}

\subsection{Validity and Usability Assessment}

\begin{figure}[htp]
\centering
\setlength{\abovecaptionskip}{0.cm}
\setlength{\belowcaptionskip}{-0.cm}
\subfigure[ ]{
\begin{minipage}[t]{0.24\linewidth}
  \includegraphics[width=\textwidth]{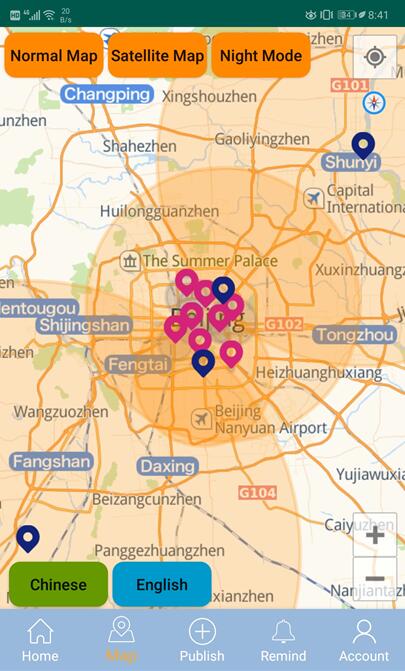}
\end{minipage}%
}%
\subfigure[ ]{
\begin{minipage}[t]{0.24\linewidth}
  \includegraphics[width=\textwidth]{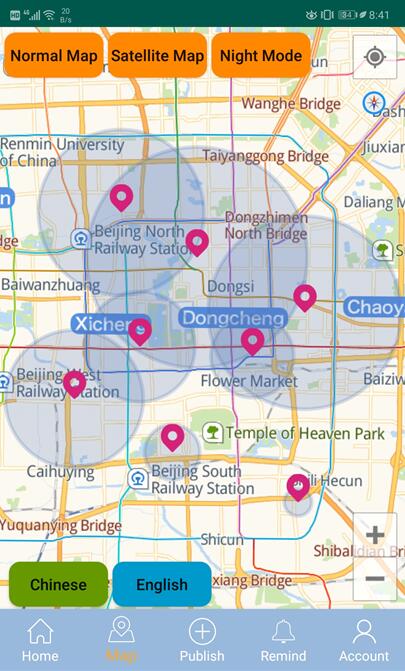}
\end{minipage}%
}%
\subfigure[ ]{
\begin{minipage}[t]{0.24\linewidth}
  \centering
  \includegraphics[width=\textwidth]{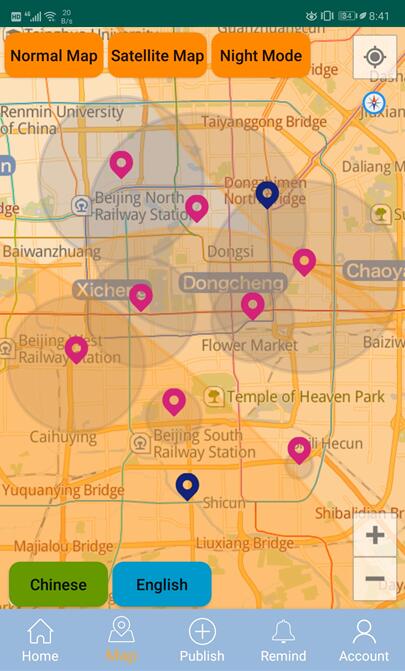}
\end{minipage}%
}%
\subfigure[ ]{
\begin{minipage}[t]{0.24\linewidth}
  \includegraphics[width=\textwidth]{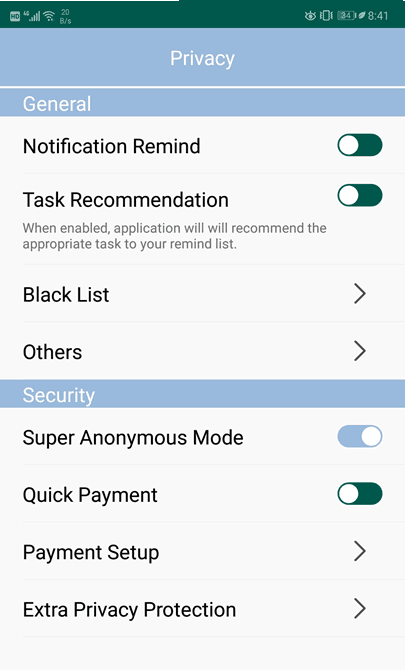}
\end{minipage}%
}%
\centering
\caption{Validity assessment interfaces. (a) task is randomly assigned, orange circles are tasks release geographic scope; (b) blue circles are the effect of using a location-based task assignment algorithm; (c) comparison of the effects of two methods; (d) choose the super privacy protection mode.}
\label{figx3}
\end{figure}

We demonstrate the validity and usability by analyzing and validating test $f_2$ and $f_3$. If the performance is improved after calling CrowdAPI, that is, using $M_2$ mode, which means that CrowdOS is indeed valid. As shown in Fig. \ref{figx3}, we show the validation verification interfaces.

Usability is reflected in the shortening of development time after calling CrowdAPI. In Fig. \ref{figx2} (a), the average time consumption of $f_2$ and $f_3$ using $M_2$ are reduced by 65.4$\%$ and 88.7$\%$ compared to using $M_1$.

In addition, with the extension of algorithm libraries, the advantages of CrowdOS are highlighted. Using it not only can greatly reduce the development time of each functional module, but also improve the overall visualization effect and program readability.

\subsection{Optimized Result Quality Assessment}

In the simulation environment, we evaluate the correction and optimization capabilities of \emph{TRO} framework joint application interface.

\begin{figure}[htbp]
\centering
\setlength{\abovecaptionskip}{0.cm}
\setlength{\belowcaptionskip}{-0.cm}
\subfigure[ ]{
\begin{minipage}[t]{0.24\linewidth}
  \includegraphics[width=\textwidth]{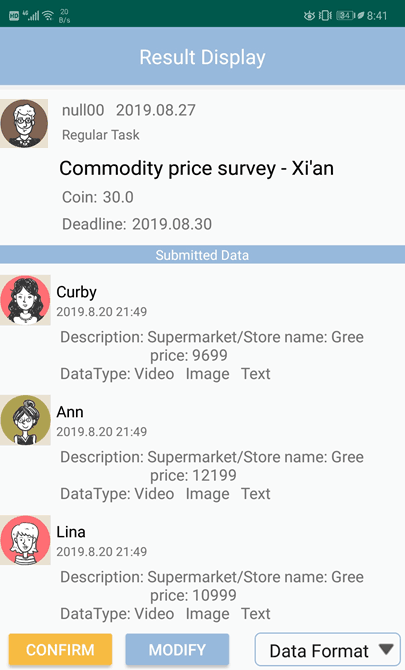}
\end{minipage}%
}%
\subfigure[ ]{
\begin{minipage}[t]{0.24\linewidth}
  \includegraphics[width=\textwidth]{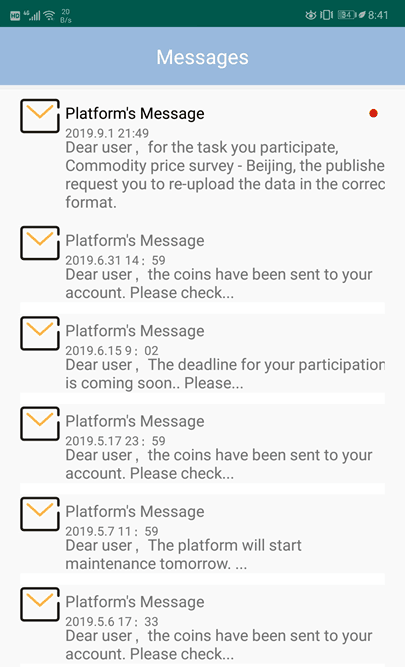}
\end{minipage}%
}%
\subfigure[ ]{
\begin{minipage}[t]{0.50\linewidth}
  \centering
  \includegraphics[width=\textwidth]{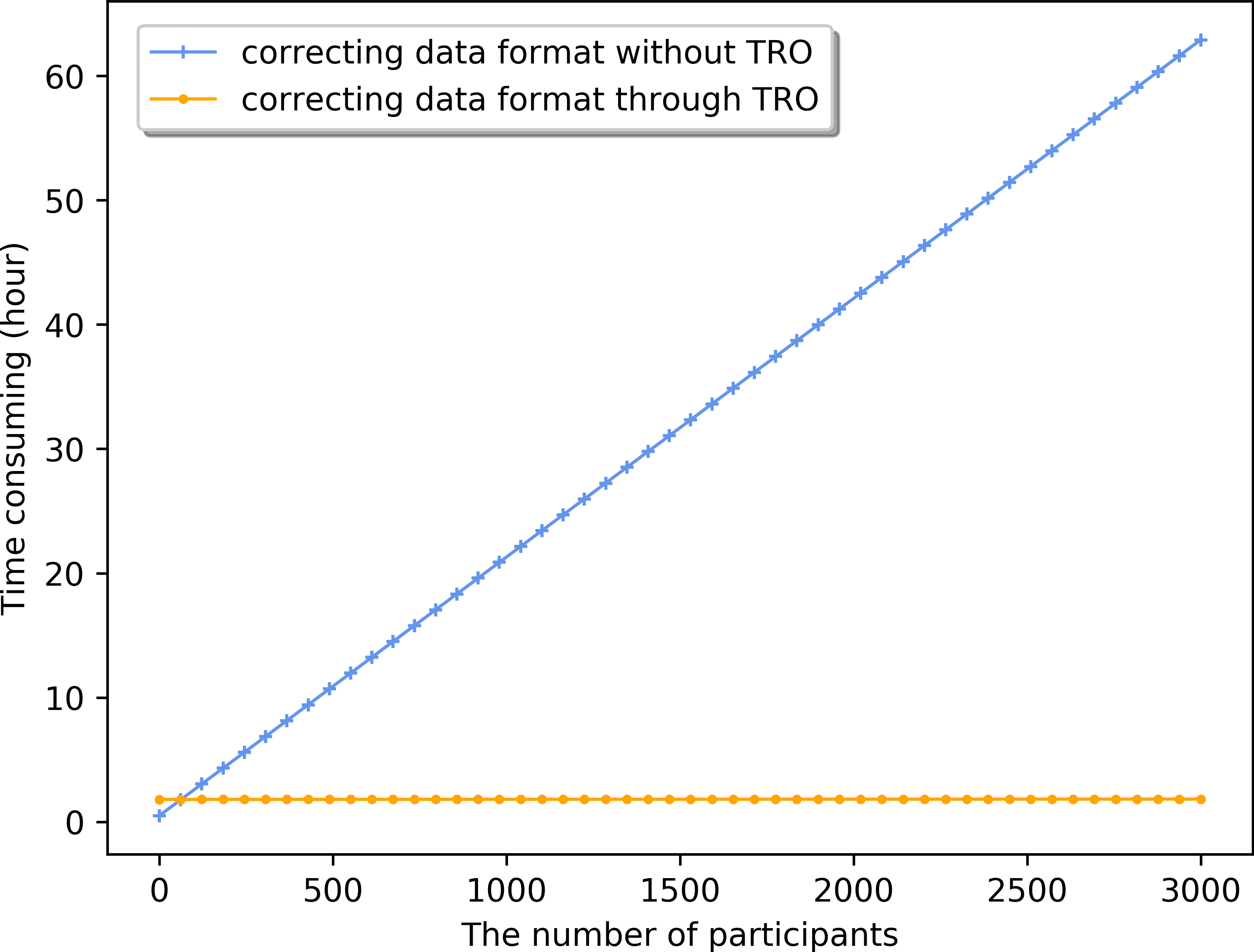}
\end{minipage}%
}%
\centering
\caption{Optimization methods and time consumption comparison. (a) data format correction request interface; (b) participant receives a correction reminder message; (c) comparison of time consumption curves for two optimization methods.}
\label{figx5}
\end{figure}

{\it Data format error problem}. For the collected data $D(n*V)$, where $n$ is the participant number and $V$ is the amount of data contributed by each participant. The time spent on correcting data format through \emph{TRO} framework is $Ti_{B}=Ti_{\zeta_B}+Ti_{\eta_B}$, where $Ti_{\zeta_B}$ is the time of interface operation ($Ti_{\zeta_B}<6 min$), the interface is shown in Fig. \ref{figx5} (a). $Ti_{\eta_B}$ is the time spent by each participant to correct format and resubmit data, $Ti_{\eta_B}\propto{V}$. Without \emph{TRO}, the time needed to correct all data format by publisher is $Ti_A$. $Ti_{A}=Ti_{\zeta_A}+Ti_{\eta_A}$, where $Ti_{\zeta_A}$ is the preparation time before correcting data format ($Ti_{\zeta_A}<30 min$), $Ti_{\eta_A}$ is the time spent on processing data formats, $Ti_{\eta_A}\propto{n*V}$.

\begin{small}
\begin{equation} \label{eqx3}
Ti_{A,B}=\begin{cases}
Ti_{\zeta_A}+Ti_{\eta_A},&Ti_{\eta_A}\propto{n*V} \\
Ti_{\zeta_B}+Ti_{\eta_B},&Ti_{\eta_B}\propto{V} \\
\end{cases}
\end{equation}
\end{small}

Eq. (\ref{eqx3}) shows the time consumed by the two optimization methods to deal with data format problems. Fig. \ref{figx5} (c) shows the distance between $A$ and $B$ curve varies with the amount of data $D(n*V)$ .

The relevance between results and task demands is reflected in multiple ways, such as the request is video but the participants submitted  images, or the actual location of the uploaded data does not match the location in the task request. Our optimization mechanism re-screens users with high reputation and update task characteristics based on specific information such as geographic location, then re-publish the task or feedback to the original participants, as shown in Fig. \ref{figx5} (b).

 \emph{TRO} not only avoids the energy consumption caused by large amount processing, but also applies to various types of tasks. The framework uses the idea of segmentation and integration to properly hand over the work of different phases to people or machines. As shown in Fig. \ref{figx5} (c), compared to other optimization methods, the time consumption of \emph{TRO} is relatively stable and does not increase significantly as the number of participants increases. Meanwhile, many optimization problems are more suitable to be solved through \emph{TRO}, and the resource consumption can be greatly reduced in comparison to pure machine optimization.

\subsection{Performance, Load and Stress Testing}

We conducted an overall test of both the Sensing-end and Server-end of \emph{WeSense} from the following two aspects.

\begin{figure}[htbp]
\centering
\setlength{\abovecaptionskip}{0.cm}
\setlength{\belowcaptionskip}{-0.cm}

{\begin{minipage}[t]{0.482\linewidth}
  \includegraphics[width=\textwidth]{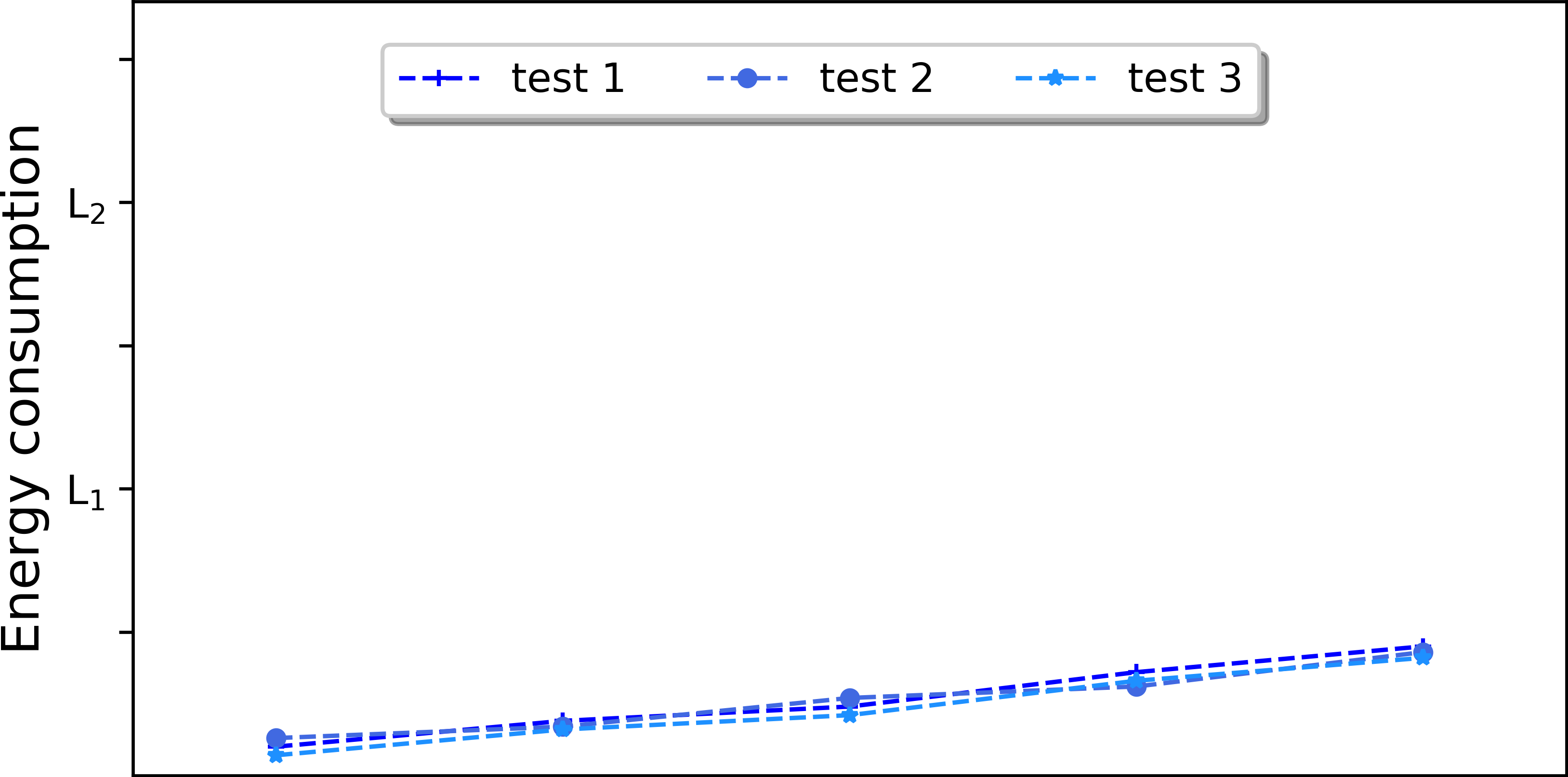}
\end{minipage}}%
\vspace{0.1cm}
\begin{minipage}[t]{0.482\linewidth}
  \includegraphics[width=\textwidth]{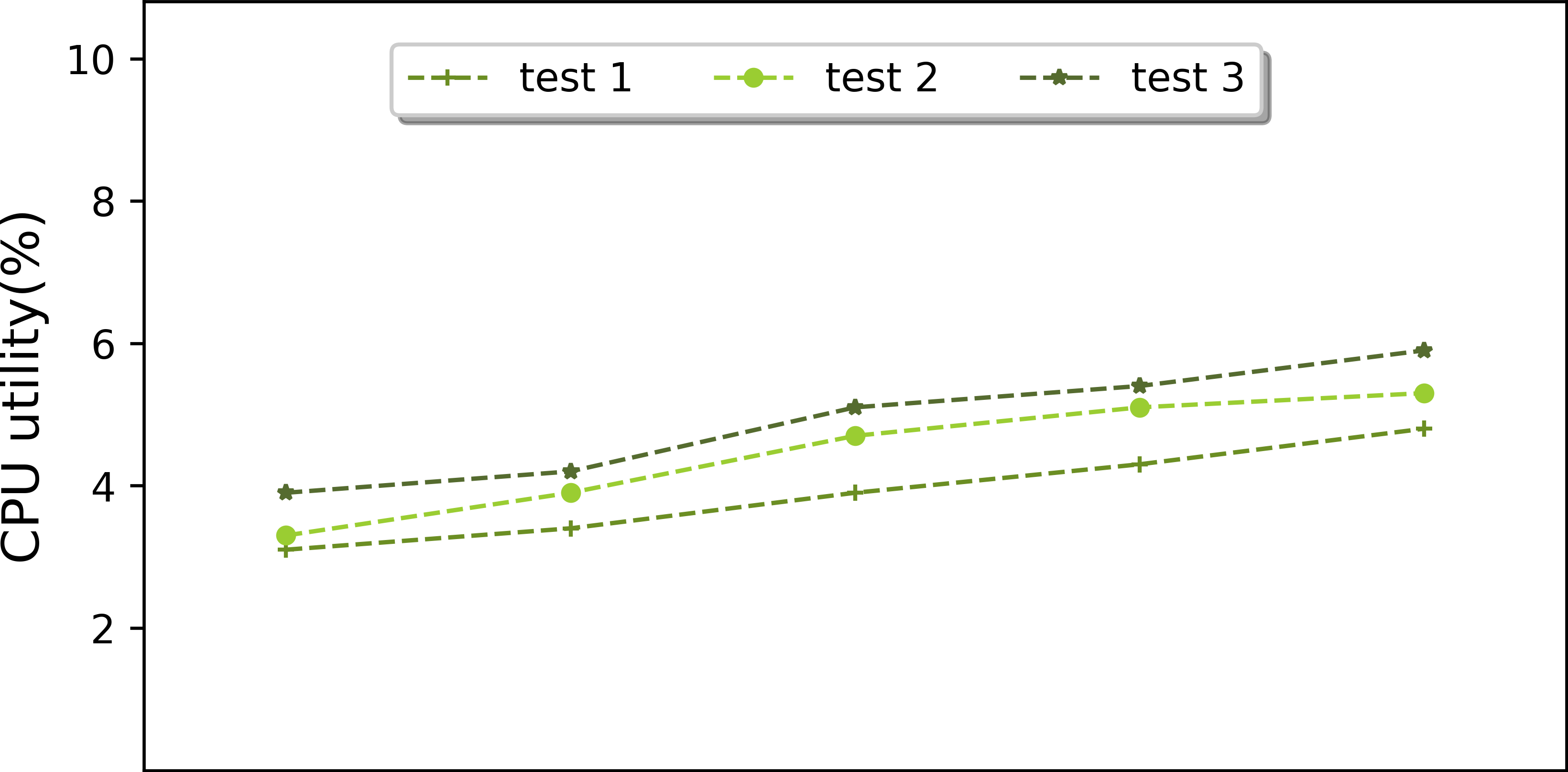}
\end{minipage}%

\begin{minipage}[t]{0.484\linewidth}
  \includegraphics[width=\textwidth]{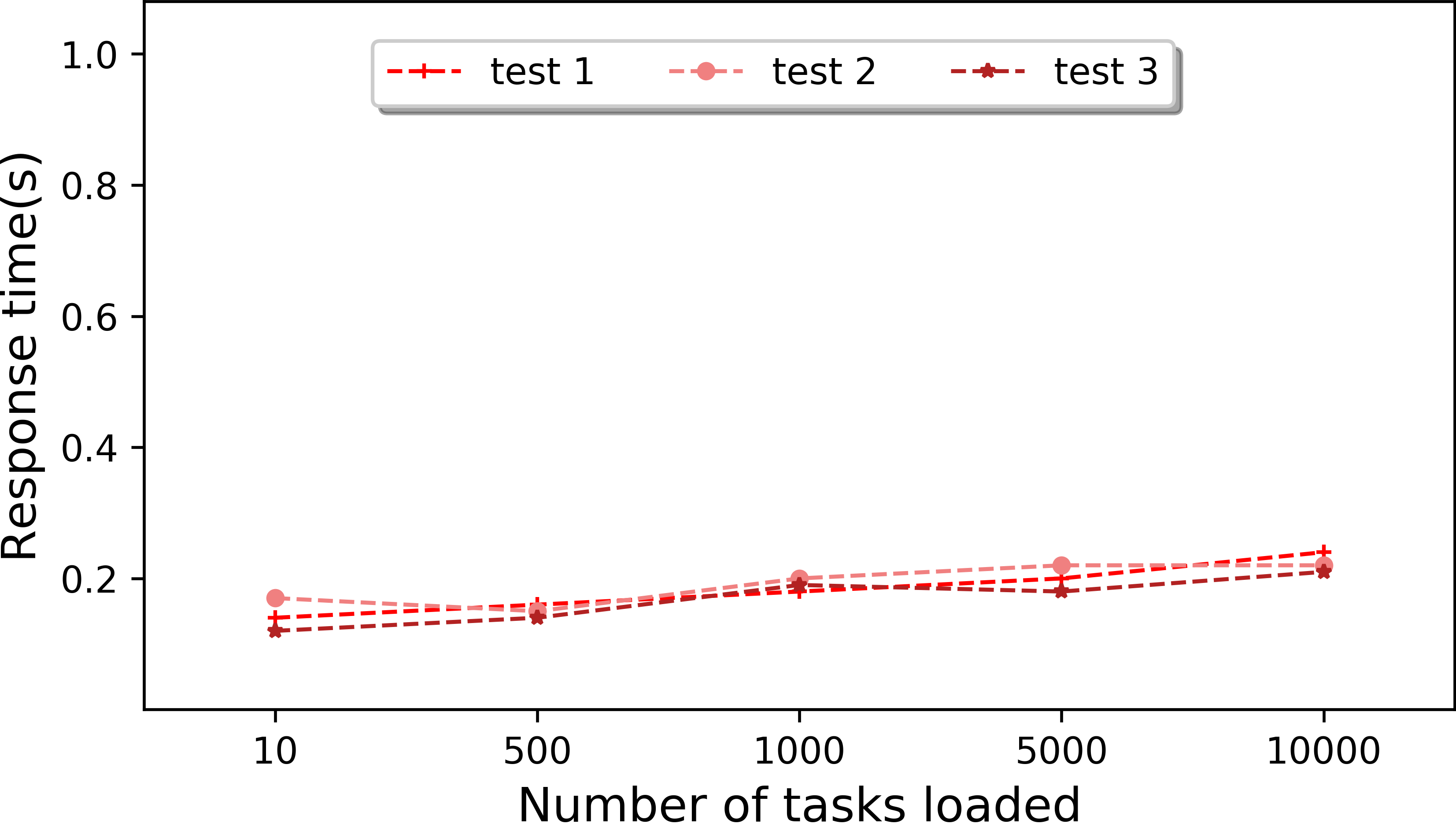}
\end{minipage}%
\vspace{0.1cm}
\begin{minipage}[t]{0.482\linewidth}
  \includegraphics[width=\textwidth]{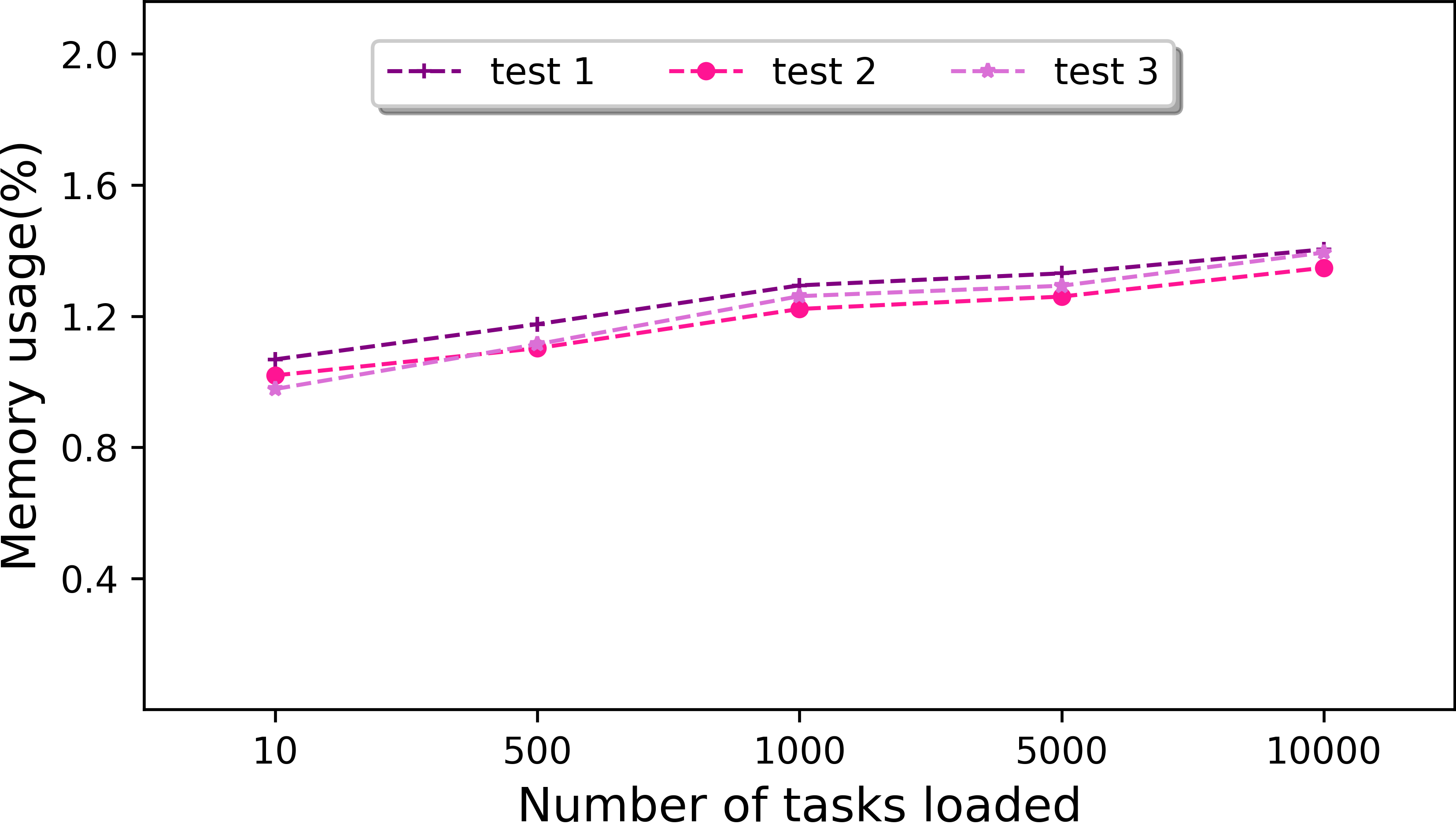}
\end{minipage}%
\centering
\caption{Performance and load testing. (We only showed three test results for each item.)}
\label{figz1}
\end{figure}

\emph{Performance and load testing}. We loaded different scales of tasks in turn and measured the system response time, CPU and memory usage, energy consumption in Sensing-end. We ran each test ten times and use profile performance analyzer of Android Studio to monitor data in real time when the application is running.  
The results are shown in Fig. \ref{figz1}. Despite the increasing number of tasks, the system response time is basically within 0.22s, and the CPU and memory usage are also maintained in 3\%-6\% and 0.87\%-1.14\% range. Energy consumption is basically kept below the minimum level ($L_{1}$: Light). This shows that the system scales well. 

\emph{Stability and stress testing}. We combine two methods for the testing. First, the server runs continuously for 7x24 hours. During the period, the number and contents of tasks are updated through mobile devices. We continuously observe and record the output logs of the sensor-end and server-end, and there is no abnormality such as crash or software error. 
Second, we use the software \emph{Android SDK Monkey} to perform stability and stress testing on \emph{WeSense} after setting up the test environment. For instance, we sent \emph{monkey -p com.hills.WeSense -v -v 1000} to request executing 1000 Random Command Events (RCE), such as Map keys, Home keys. We recorded the number of occurrences of CRASH and ANR (Application Not Responding). CRASH refers to the situation where the program stops or exits abnormally when an application error occurs. ANR means that when the Android system detects that the application does not respond to input events within 5 seconds or the broadcast does not execute within 10 seconds, it throws an unresponsive prompt. The test results are shown in Table \ref{table7}. Combining the above two test results, we can see that the system can run stably and efficiently under different pressure conditions.

\begin{table}[!thbp]\footnotesize
\newcommand{\tabincell}[2]{\begin{tabular}{@{}#1@{}}#2\end{tabular}}
\caption{Stability and stress testing}
\label{table7}
\centering
\vspace{-0.25cm}
\begin{tabular}{p{0.85cm}p{0.8cm}p{0.8cm}p{0.8cm}p{0.8cm}p{0.8cm}}
\toprule
\multicolumn{1}{c}{\textbf{RCE number}} & \textbf{1000}& \textbf{5000} & \textbf{10000}& \textbf{50000}& \textbf{100000}\\
\midrule
{\tabincell{c}{CRASH times\\ANR times\\\ \ TOTAL}}&{\tabincell{c}{\ \ \ 0\\\ \ \ 0\\\ \ \ 0}}&{\tabincell{c}{\ \ \ 0\\\ \ \ 0\\\ \ \ 0}}&{\tabincell{c}{\quad0\\\quad0\\\quad0}}&{\tabincell{c}{\quad0\\\quad0\\\quad0}}&{\tabincell{c}{\quad\ 0\\\quad\ 0\\\quad\ 0}}\\
\bottomrule
\end{tabular}
\end{table}

\section{Discussion and Future Work}

In this new research field, there are still many issues to be further investigated and improved. We discuss some potential limitations in our work next. 

First of all, we consider the balance between complexity and maintainability of system. The OS kernel is sophisticated and intricate. In addition to the important frameworks such as \emph{TRUS}, \emph{IRM}, and \emph{DFHMI}, there are also a variety of functional modules interacting with each other. From systematical point of view, these modules work together through complex mechanisms to implement system functions. However, considering the principles of software engineering, we put the functions into separate and independent modules at the beginning, which interacts through interfaces to reduce coupling. The system further upgrades and manages each algorithm, function module or interface separately, which lowers difficulty in maintenance.

Second, we discuss the extension of result quality optimization methods. \emph{DFHMI} is  a fundamental and versatile framework, but it does not include specific data optimization algorithms for tasks. However, we provide libraries and interfaces for specific optimization methods. System Plugins in Fig. \ref{fig2} are extensible and not fully enumerated, such as Multiple Types of Data Processing (\emph{MTDP}) plugin. Developers and contributors can package and upload  relevant data processing algorithms to the \emph{MTDP} library according to the interface protocol, and then it can be called in applications. For example, if a developer builds a water quality monitoring application through CrowdOS, he can call the water quality data optimization package in the library of \emph{MTDP} plugin to help users get higher quality results.

Third, how to maintain system stability is challenging in the presense of a large number of machine learning models that can be updated online. For example, \emph{SAS}, \emph{RDT}, and \emph{SDIM} all have dynamic self-update capabilities that can be adjusted as the task size increases. This  can cause operating system stability issues. To solve this problem, we added constraints and fallback functions to the correlation model to prevent them from losing control. Once the system detects an anomaly, the problematic model will automatically roll back to the previous version. The modules in the system are maintained separately, interact through the interface, have low coupling characteristics, and have a clear version management module, which helps to locate the cause and update in time.

Fourth, CrowdOS is a comprehensive architecture that incorporates a lifelong learning philosophy. There are massive diversification tasks in \emph{CAP}. Further, mechanisms such as \emph{KBM} help to store and manage long-term and short-term knowledge that is mined during task execution. The knowledge is further integrated into the OS kernel to re-optimize the system processing flow and improve operational performance. In fact, the system migrates the accumulated knowledge to new tasks and updates itself over and over again. Due to the complexity of the system and lifelong learning characteristics, such as knowledge accumulation, strategy update, model migration, and deep reasoning, we need to conduct in-depth exploration and large-scale continuous testing to further evaluate and improve the kernel mechanism.

Last, we discuss how to use the system. Current CrowdOS users mainly include three types, system kernel maintainers, functional component developers, and third-party application developers. The operating system provides a rich set of functional components and mechanisms. When developers build \emph{CAP} based on the OS, in addition to using the minimal kernel module, they can also customize other libraries and modules according to individual needs. \emph{WeSense} mentioned in Section 8 can be downloaded from the project website \cite{crowdos.cn}. It is a comprehensive crowdsourcing platform where users can conduct task transactions. 

\section{Conclusion}

In this paper, we present a ubiquitous operating system, CrowdOS, to solve the problem of the lack of a unified architecture for existing $\mathbb{CAP}$ and the incompatibility of algorithms or modules in related research. We elaborated on the kernel architecture and focused on implementing three of the core frameworks: \emph{TRAF} establishes a bridge between tasks and OS kernel through \emph{TRG}, and then adaptively selects  reasonable allocation strategies for heterogeneous tasks; \emph{IRM} abstracts heterogeneous physical and virtual resources in the system and provides them with unified software definition and management; \emph {DFHMI} is designed to quantify and optimize the quality of results via quality assessment and shallow-deep inference mechanisms, as well as strategies that integrate specific quality issues. Through the analysis of the development process of \emph{WeSense}, we evaluated the correctness of CrowdOS and the effectiveness of kernel modules, as well as the overall development efficiency, and also compared the optimization speed and energy consumption of the results before and after using \emph {TRO}.

\ifCLASSOPTIONcompsoc
  \section*{Acknowledgments}
  Thanks to the members of CrowdOS project team and volunteers. This work was supported in part by the National Science Fund for Distinguished Young Scholars (No. 61725205), the National Key R\&D Program of China (No. 2018YFB2100800), and the National Natural Science Foundation of China (No. 61772428). 
\else
  \section*{Acknowledgment}
\fi

\ifCLASSOPTIONcaptionsoff
  \newpage
\fi


\bibliographystyle{IEEEtran}
\bibliography{./IEEEabrv}

\begin{IEEEbiography}[{\includegraphics[width=0.95in,height=1.25in,clip]{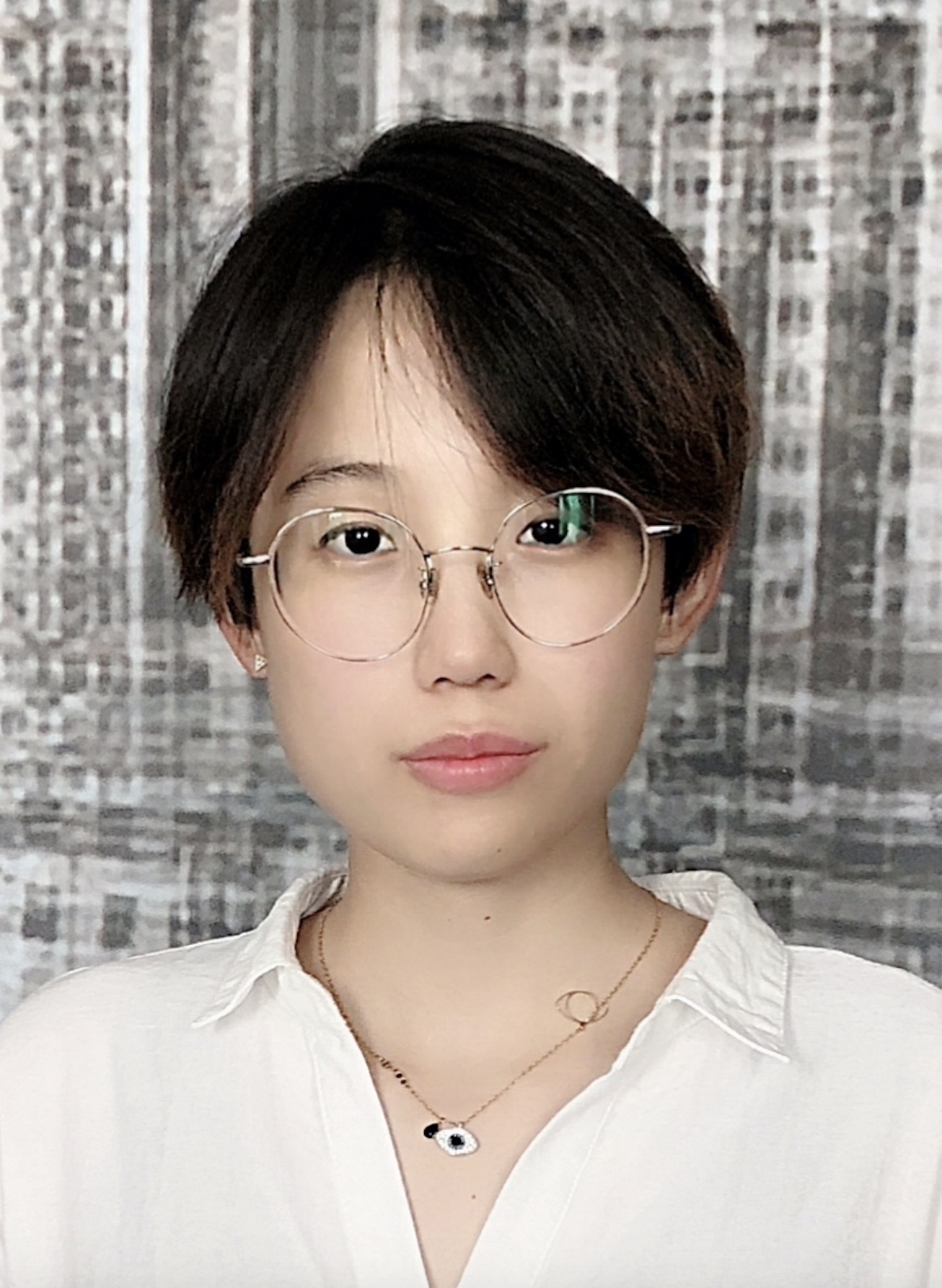}}]{Yimeng Liu}
received the MEng degree in computer science from University of Chinese Academy of Sciences, Beijing, China, in 2017. She is currently working toward the PhD degree in the Department of Computer Science, Northwestern Polytechnical University, China. Her research interests include ubiquitous computing, ubiquitous operating system, and artificial intelligence.
\end{IEEEbiography}

\begin{IEEEbiography}[{\includegraphics[width=0.95in,height=1.25in,clip]{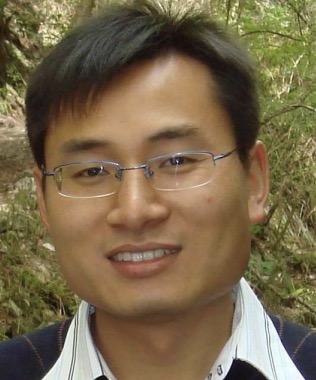}}]{Zhiwen Yu}
received the Ph.D. degree in computer science from Northwestern Polytechnical University, Xi'an, China, in 2006. He is currently a Professor and the Vice-Dean of the School of Computer Science, Northwestern Polytechnical University, Xi'an, China. He was an Alexander Von Humboldt Fellow with Mannheim University, Germany, and a Research Fellow with Kyoto University, Kyoto, Japan. His research interests include ubiquitous computing and HCI.
\end{IEEEbiography}

\begin{IEEEbiography}[{\includegraphics[width=0.95in,height=1.25in,clip]{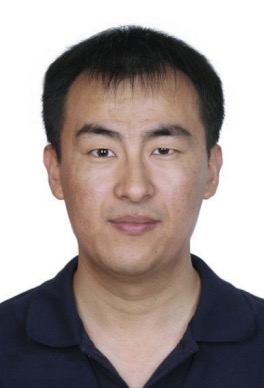}}]{Bin Guo}
received the Ph.D. degree in computer science from Keio University, Minato, Japan, in 2009, He was a Postdoctoral Researcher with the Institut TELECOM SudParis, Essonne, France. He is currently a Professor with Northwestern Polytechnical University, Xi'an, China. His research interests include ubiquitous computing, mobile crowd sensing, and HCI.
\end{IEEEbiography}

\begin{IEEEbiography}[{\includegraphics[width=0.95in,height=1.25in,clip]{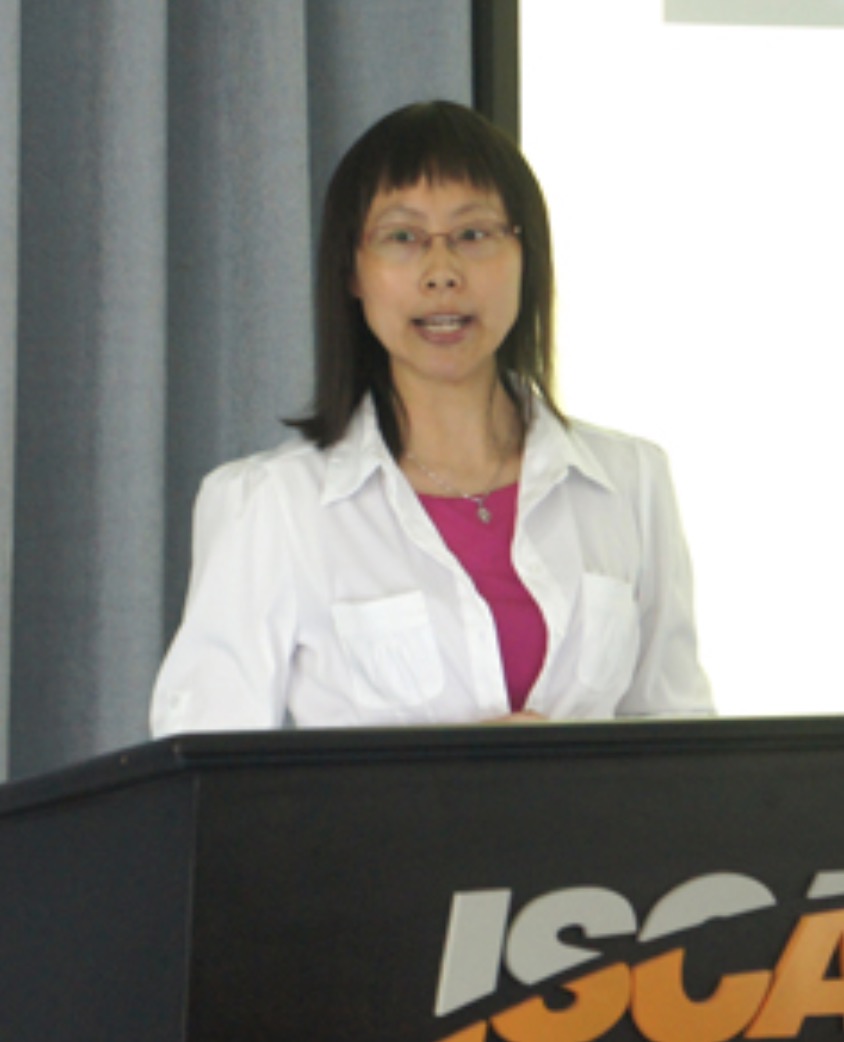}}]{Qi Han} 
is a Professor in the Department of Computer Science from Colorado School of Mines, US. She obtained her Ph.D. degree in Computer Science from the University of California-Irvine in August 2005. Her research interests include mobile crowd sensing and ubiquitous computing.
\end{IEEEbiography}

\begin{IEEEbiography}[{\includegraphics[width=0.95in,height=1.25in,clip]{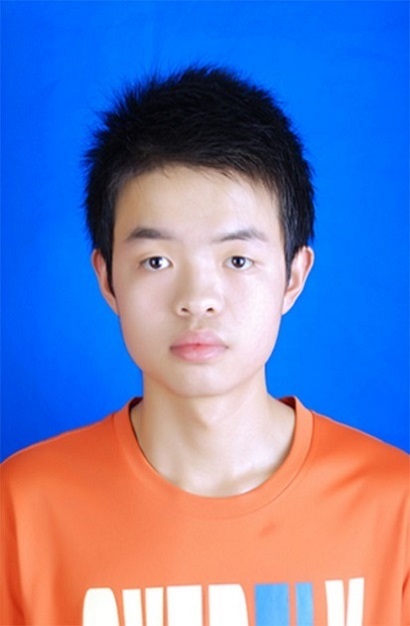}}]{Jiangbin Su} 
is a graduate student in the Department of Computer Science and Technology from Northwestern Polytechnic University, China. He obtabined his Bachelor degree in Computer Science and Technology from northwestern polytechnic university in June 2019. His research insterests include mobile crowd sensing and big data.
\end{IEEEbiography}

\begin{IEEEbiography}[{\includegraphics[width=0.95in,height=1.25in,clip]{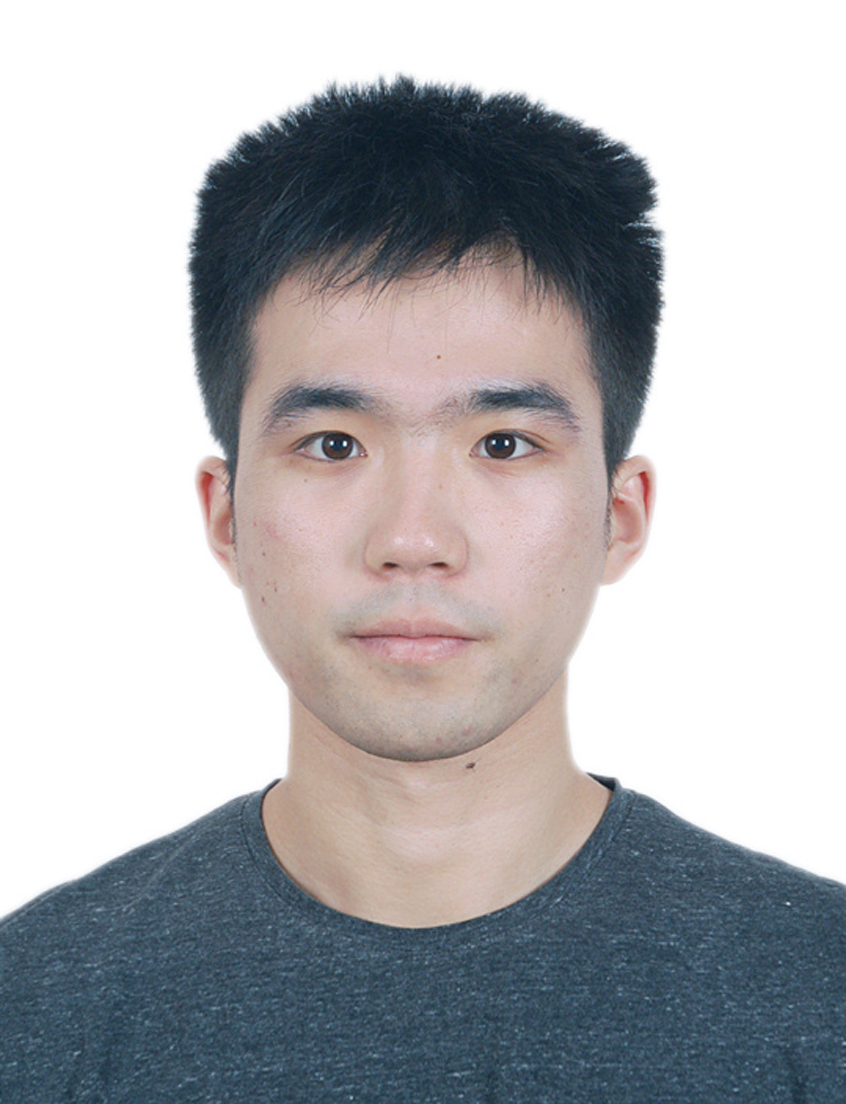}}]{Jiahao Liao} 
is a graduate student in the Department of Software Engineering from Northwestern Polytechnic University, China. He obtabined his Bachelor degree in Software Engineering from Northwestern Polytechnic University in June 2019. His research insterests include mobile crowd sensing and software engineer.
\end{IEEEbiography}







\end{document}